\documentclass{article}

     \usepackage[preprint]{neurips_2022}

\usepackage[utf8]{inputenc} %
\usepackage[T1]{fontenc}    %
\usepackage{hyperref}       %
\usepackage{url}            %
\usepackage{booktabs}       %
\usepackage{amsfonts}       %
\usepackage{nicefrac}       %
\usepackage{microtype}      %
\usepackage{xcolor}         %
\usepackage{bm}
\usepackage{amsmath}

\usepackage{graphicx}
\usepackage{verbatim}
\usepackage{float}
\usepackage{amsmath}
\usepackage{breqn}

\title{Neural tangent kernel analysis of PINN for advection-diffusion equation}

\author{M. H. Saadat, B. Gjorgiev,  L. Das and
   G. Sansavini \\ %
   Reliability and Risk Engineering Laboratory, Institute of Energy and Process Engineering,\\ Department of Mechanical and Process Engineering, ETH Zurich, 8092 Zurich, Switzerland \\
   \\
   \texttt{gsansavini@ethz.ch} \\
}

\begin{document}

\maketitle

\begin{abstract}
  Physics-informed neural networks (PINNs) numerically approximate the solution of a partial differential equation (PDE) 
  by incorporating the residual of the PDE along with its initial/boundary conditions into the loss function. In spite of their partial success, PINNs are known to struggle
  even in simple cases where the closed-form analytical solution is available. 
  In order to better understand the learning mechanism of PINNs, this work focuses on a systematic analysis of PINNs for the linear advection-diffusion equation (LAD) using the Neural Tangent Kernel (NTK) theory. Thanks to the NTK analysis, the effects of the advection speed/diffusion parameter on the training dynamics of PINNs are studied and clarified. We show that the training difficulty of PINNs is a result of 1) the so-called 'spectral bias', which leads to difficulty in learning high-frequency behaviours; and 2) convergence rate disparity between different loss components that results in training failure. The latter occurs even in the cases where the solution of the underlying PDE does not exhibit high-frequency behaviour. Furthermore, we observe that this training difficulty manifests itself, to some extent, differently in advection-dominated and diffusion-dominated regimes. Different strategies to address these issues are also discussed. In particular, it is demonstrated that periodic activation functions can be used to partly resolve the spectral bias issue.
  
\end{abstract}

\section{Introduction}
Developing accurate and efficient numerical schemes for solving partial differential equations (PDEs) is an active research field with enormous applications in different scientific and engineering disciplines. Classical numerical methods, such as finite difference, finite volume or finite element schemes, rely on discretizing the governing equations over an underlying grid points and marching forward in time, and have been successfully applied to a wide range of applications from fluid dynamics and quantum mechanics to neuroscience and finance.

In addition to above mentioned established numerical methods, neural networks (NNs) have recently received attention as an alternative approach to conventional PDE solvers~\cite{yu2018deep, sirignano2018dgm}. In particular, the first realization of the physics-informed neural network (PINN)~\cite{raissi2019physics} showed promising results for solving various linear and non-linear PDEs. PINN approximates the solution of a PDE by turning it into an unsupervised optimization problem with the constraints being the residual of the PDE under consideration and its initial and boundary conditions. The major attractiveness of PINN, however, lies in the computation of derivatives which is done through the automatic differentiation (AD) technique during back-propagation. This results in a mesh-free algorithm that opens up the possibility of accurately solving complex and computationally demanding PDEs including high-dimensional PDEs \cite{han2018solving} or Integral-differential equations~\cite{yuan2022pinn}. The pioneering work of~\cite{raissi2019physics} has inspired various improved PINN models which extend the operating domain of the original model~\cite{raissi2019physics} and are to some extent capable of solving PDEs in relevant practical applications ~\cite{lou2021physics, jagtap2020conservative, jin2021nsfnets, gao2022physics, kashefi2022physics, meng2020ppinn, raissi2020hidden, mao2020physics}. See~\cite{cuomo2022scientific} for a full review of PINNs development. 

The original realization of PINN~\cite{raissi2019physics}, however, struggles in dealing with PDEs which exhibit high-frequency or multi-scale behaviour, even if the underlying PDE has a closed-form analytical solution. It has been shown that in those scenarios the PINN fails to learn the correct solution, no matter how large the architecture is \cite{krishnapriyan2021characterizing}. Different studies have attempted to identify the failure modes of PINN and to understand its learning mechanism \cite{krishnapriyan2021characterizing, wang2021understanding, mishra2022estimates, de2021error}. In particular, PINN failure has been attributed to the complexity of the loss function in challenging physical regimes~\cite{krishnapriyan2021characterizing} or to the unbalanced gradients of different loss terms during training, which results in numerical stiffness and unbalanced back-propagation via Gradient Descnet (GD) \cite{wang2021understanding}. Yet, the most rigorous mathematical explanation of the PINNs failure has been presented in~\cite{wang2022and} based on the so-called Neural Tangent Kernel (NTK) theory~\cite{jacot2018neural}. 

NTK is a recent theory which studies fully-connected NNs in the infinite-width limit, i.e, in the limit as the number of neurons in the hidden layers tend to infinity \cite{jacot2018neural, arora2019exact}. It has been shown that in this limit, NNs behave like a linear model obtained from the first-order Taylor expansion of the NN around its initial parameters \cite{jacot2018neural, arora2019exact, lee2019wide}. In other words, according to NTK theory, fully-connected infinite width NNs trained by the GD algorithm behave like a kernel regression model with a deterministic kernel called NTK \cite{jacot2018neural, arora2019exact, lee2019wide}. The NTK theory has clarified different features of the NN training, including the success of over-parameterized networks. Another feature worth mentioning, is the so-called 'spectral-bias' or 'F-principle' \cite{rahaman2019spectral, xu2019frequency}, which describe the tendency of NNs to be biased towards low-frequency solutions.

Wang et al.~\cite{wang2022and} derived the NTK for PINNs and proved its convergence to a deterministic kernel, provided that the network is wide and the learning rate is small enough. Based on that, they showed that, similar to NNs, PINNs also suffer from the 'spectral bias'~\cite{wang2022and, wang2021eigenvector}. Furthermore, they demonstrated that the convergence rate discrepancy between different components in the loss function can lead to PINN failure~\cite{wang2022and, wang2021eigenvector}. 

Regardless of limitations of the NTK theory, the framework presented by~\cite{wang2022and} provides an exceptional step forward towards analysing and understanding the behaviour of PINNs in dealing with different class of PDEs. Building on this work~\cite{wang2022and}, this paper aims at (1) apply the NTK theory to PINNs for the linear advetion-diffusion (LAD) equation, (2) clarify the exact mechanisms behind PINNs failure in advection-dominated and diffusion-dominated regimes, and (3) exploring the effectiveness of different training strategies in the view of NTK theory, and particularly the role of activation function on training dynamics. All theoretical results are supplemented by numerical experiments. We choose to study the advection-diffusion equation, as it describes two fundamental physical processes with enormous applications from fluid dynamics to biological systems and financial mathematics. In addition, there is a closed-form analytical solution, as well as a mature literature on different numerical methods for dealing with LAD equations. As a result, studying PINNs for this class of PDEs would provide valuable insights into application of PINNs for more challenging and complicated PDEs.

The remainder of the paper is organized as follows:
for the sake of completeness, Section~\ref{sec:methodology} briefly presents the main steps of deriving NTK for PINNs. Section~\ref{sec:numExper}, studies the PINN for the LAD and clarifies the impact of the PDE parameters on the training dynamics of PINN through NTK analysis backed by numerical experiments. Furthermore, the effectiveness of different improved learning strategies
are investigated. Particularly, the role of activation function is studied and analyzed. Finally, Section~\ref{sec:discussion} discusses our findings and provides conclusions.

\section{Methodology}
\label{sec:methodology}
In this section, we briefly summarize the neural tangent kernel derivation of PINNs for the linear advection-diffusion equations following~\cite{wang2022and}. Interested readers are referred to~\cite{wang2022and, wang2021eigenvector} for the full derivation and proofs of the NTK theory for PINNs. 

Let us, for simplicity, consider the following scalar advection-diffusion equation,
\begin{align}
    {u}_t + a u_x - \epsilon u_{xx} &= 0, {x} \in \Omega,  \label{eq:LDA} \\ \nonumber
    {u}({x},0) &= f(x), {x} \in \Omega, \\ \nonumber
    {u}({x},t) &= g(x), {x} \in \partial \Omega,
\end{align}
where ${u}({x},t): \Omega \times [0,T] \rightarrow \mathbb{R}$ denotes the unknown solution in one-dimensional physical space $\Omega$, $a$ is the advection speed, and $\epsilon$ represents the diffusion parameter. For the sake of notational simplicity, we can consider time $t$ as an additional dimension in $x$, i.e. $\bm{x} = (x,t)$.

Following the original formulation of PINN~\cite{raissi2019physics}, we approximate the solution ${u}({x},t) = u(\bm{x})$ with a deep neural network ${u}(\bm{x},\bm{\theta})$, where $\bm{\theta} \in \mathbb{R}^{n_{par}}$ is a collection of all network parameters including weights and biases, and by considering the following loss function,
\begin{align}
    \mathcal{L} &= \mathcal{L}_r + \mathcal{L}_b,
    \label{eq:Loss}
\end{align}
where 
\begin{align}
    \mathcal{L}_r &=  \frac{1}{2} \sum_{i=1}^{N_r} \| {u}_t(\bm{x}_r^i,\bm{\theta}) + a {u}_x(\bm{x}_r^i,\bm{\theta}) - \epsilon u_{xx}(\bm{x}_r^i,\bm{\theta}) \|^2 = \frac{1}{2} \sum_{i=1}^{N_r} \| \mathcal{N} u(\bm{x}_r^i,\bm{\theta}) \|^2, \label{eq:Loss_r}  \\
    \mathcal{L}_b &= \frac{1}{2} \sum_{i=1}^{N_b} \| {u}(\bm{x}_b^i,\bm{\theta}) - g(\bm{x}_b^i) \|^2.
    \label{eq:Loss_b}
\end{align}
Here, $\mathcal{N}$ represents the differential operator on the LHS of \ref{eq:LDA}, $\mathcal{L}_r$ denotes the loss over residual of the PDE (\ref{eq:LDA}) at $N_r$ collocation points sampled from inside of the domain $\{\bm{x}_r^i\}_{i=1}^{N_r}$, and $\mathcal{L}_b$ is the loss associated with initial/boundary conditions evaluated at $N_b$ initial/boundary points, $\{\bm{x}_b^i\}_{i=1}^{N_b}$.

Learning the network parameters $\bm{\theta}$, 
implies %
minimization of the loss function (\ref{eq:Loss}) through the gradient descent (GD) algorithm. Assuming an infinitesimal learning rate, GD can be written in terms of the so-called continuous time gradient flow differential equation \cite{wang2022and},
\begin{align}
    \frac{d \bm{\theta}}{d t} = - \nabla_{\bm{\theta}} \mathcal{L}(\bm{\theta}). 
\end{align}
Applying the gradient leads to,
\begin{align}
     \frac{d \bm{\theta}}{d t} = -\left[ \sum_{i=1}^{N_r} \left( \mathcal{N} u(\bm{x}_r^i,\bm{\theta}(t)) \right) \frac{\partial \mathcal{N} u(\bm{x}_r^i,\bm{\theta}(t))}{\partial \bm\theta} + \sum_{i=1}^{N_b} \left( {u}(\bm{x}_b^i,\bm{\theta}(t)) - g(\bm{x}_b^i) \right) \frac{\partial {u}(\bm{x}_b^i,\bm{\theta}(t))}{\partial \bm\theta} \right]. \label{eq:dtheta/dt}
\end{align}
Using the chain rule, we can write for  $0 \leq k \leq N_r$ and $0 \leq j \leq N_b$,
\begin{align}
    \frac{d  \mathcal{N}u(\bm{x}_r^k,\bm{\theta}(t))}{dt} &= \frac{d  \mathcal{N}u(\bm{x}_r^k,\bm{\theta}(t))}{d \bm\theta}^T \cdot \frac{d \bm\theta}{dt}, \label{eq:dN/dt} \\
    \frac{d  u(\bm{x}_b^j,\bm{\theta}(t))}{dt} &= \frac{d  u(\bm{x}_b^j,\bm{\theta}(t))}{d \bm\theta}^T \cdot \frac{d \bm\theta}{dt}. \label{eq:du/dt}
\end{align}
Substituting (\ref{eq:dtheta/dt}) into (\ref{eq:dN/dt}) results in,
\begin{dmath}
    \frac{d  \mathcal{N}u(\bm{x}_r^k,\bm{\theta})}{dt} = - \left[\sum_{i=1}^{N_r} \left( \mathcal{N} u(\bm{x}_r^i,\bm{\theta}) \right) \langle \frac{d  \mathcal{N}u(\bm{x}_r^k,\bm{\theta})}{d \bm\theta} , \frac{d \mathcal{N} u(\bm{x}_r^i,\bm{\theta})}{d \bm\theta} \rangle
    +  \sum_{i=1}^{N_b} \left( {u}(\bm{x}_b^i,\bm{\theta}) - g(\bm{x}_b^i) \right) \langle \frac{d  \mathcal{N}u(\bm{x}_r^k,\bm{\theta})}{d \bm\theta} ,  \frac{d {u}(\bm{x}_b^i,\bm{\theta})}{d \bm\theta} \rangle 
     \right], \label{eq:dN/dt_}
\end{dmath}
and similarly for (\ref{eq:du/dt}), 
\begin{dmath}
    \frac{d  u(\bm{x}_b^j,\bm{\theta})}{dt} = - \left[\sum_{i=1}^{N_r} \left( \mathcal{N} u(\bm{x}_r^i,\bm{\theta}) \right) \langle \frac{d  u(\bm{x}_b^j,\bm{\theta})}{d \bm\theta} , \frac{d \mathcal{N} u(\bm{x}_r^i,\bm{\theta})}{d \bm\theta} \rangle 
     +  \sum_{i=1}^{N_b} \left( {u}(\bm{x}_b^i,\bm{\theta}) - g(\bm{x}_b^i) \right) \langle \frac{d  u(\bm{x}_b^j,\bm{\theta})}{d \bm\theta} ,  \frac{d {u}(\bm{x}_b^i,\bm{\theta})}{d \bm\theta} \rangle 
     \right], \label{eq:du/dt_}
\end{dmath}
where the part within $\langle  \rangle$ indicates the inner product over all neural network parameters $\bm \theta$. By defining the following kernels, 
\begin{align}
    \left(\bm K_{u}\right)_{ij}(t) &= \langle \frac{d  u(\bm{x}_b^i,\bm{\theta}(t))}{d \bm\theta} , \frac{d u(\bm{x}_b^j,\bm{\theta}(t))}{d \bm\theta} \rangle,  \\
    \left(\bm K_{ur}\right)_{ij}(t) &= \langle \frac{d  u(\bm{x}_b^i,\bm{\theta}(t))}{d \bm\theta} , \frac{d \mathcal{N} u(\bm{x}_r^j,\bm{\theta}(t))}{d \bm\theta} \rangle,  \\
    \left(\bm K_{r}\right)_{ij}(t) &= \langle \frac{d  \mathcal{N}u(\bm{x}_r^i,\bm{\theta}(t))}{d \bm\theta} , \frac{d \mathcal{N} u(\bm{x}_r^j,\bm{\theta}(t))}{d \bm\theta} \rangle, 
\end{align}
we can rewrite Eqs.~(\ref{eq:dN/dt_}) and (\ref{eq:du/dt_}) in a compact form as, 
\begin{align}
    \begin{bmatrix}
        \frac{d  u(\bm{x}_b,\bm{\theta}(t))}{dt} \\
        \frac{d  \mathcal{N}u(\bm{x}_r,\bm{\theta}(t))}{dt}
    \end{bmatrix}
    = - \bm K(t) \cdot 
    \begin{bmatrix}
         {u}(\bm{x}_b,\bm{\theta}(t)) - g(\bm{x}_b)  \\
         \mathcal{N} u(\bm{x}_r,\bm{\theta}(t)) 
    \end{bmatrix},
\end{align}
where 
\begin{align}
     \bm K(t)  = 
     \begin{bmatrix}
    \bm K_{u}(t) & \bm K_{ur}(t)\\
    \bm K_{ur}^T(t) & \bm K_{r}(t)
    \end{bmatrix},
\end{align}
is called the neural tangent kernel (NTK) of our PINN. It can be verified that for an over-parameterized PINN, the NTK is a positive semi-definite (PSD) matrix \cite{wang2022and}.%

According to the neural tangent kernel (NTK) theory, in the limit where the width of the deep neural network goes to infinity (and also under some other conditions \cite{wang2022and}), the kernel $\bm{K}(t)$ converges to a static kernel $\bm{K}^*$ and remains almost constant during the training. In most cases, the kernel can be well approximated with its value at the initialization, i.e.,
\begin{align}
    \bm K(t) \approx \bm{K}^* \approx  \bm K(0).
\end{align}
Therefore, learning dynamics can be simplified into the following system of ordinary differential equations,
\begin{align}
    \begin{bmatrix}
        \frac{d  u(\bm{x}_b,\bm{\theta})}{dt} \\
        \frac{d  \mathcal{N}u(\bm{x}_r,\bm{\theta})}{dt}
    \end{bmatrix}
    \approx - \bm K(0) \cdot 
    \begin{bmatrix}
         {u}(\bm{x}_b,\bm{\theta}) - g(\bm{x}_b)  \\
         \mathcal{N} u(\bm{x}_r,\bm{\theta}) 
    \end{bmatrix},
\end{align}
which implies that,
\begin{align}
    \begin{bmatrix}
        {u(\bm{x}_b,\bm{\theta}(t)) - g(\bm{x}_b) } \\
        {\mathcal{N}u(\bm{x}_r,\bm{\theta}(t))}
    \end{bmatrix}
    \approx e^{-\bm K(0) t} \cdot 
    \begin{bmatrix}
        {u\left(\bm{x}_b,\bm{\theta}(0) \right) - g(\bm{x}_b) } \\
        {\mathcal{N}u\left(\bm{x}_r,\bm{\theta}(0) \right)}
    \end{bmatrix}.
\end{align}
Spectral decomposition of $\bm K(0)$ as $\bm K(0) = P^T \Lambda P$, with $P$ being an orthogonal matrix whose columns are eigenvectors of $\bm K(0)$ and $\Lambda$ being a diagonal matrix with diagonal entries $\lambda_i$ corresponding to eigenvalues of $\bm K(0)$, allows us to write the solution in the following form,
\begin{align}
    \begin{bmatrix}
        {u(\bm{x}_b,\bm{\theta}(t)) - g(\bm{x}_b) } \\
        {\mathcal{N}u(\bm{x}_r,\bm{\theta}(t))}
    \end{bmatrix}
    \approx P^T e^{-\Lambda t} P \cdot 
    \begin{bmatrix}
        {u\left(\bm{x}_b,\bm{\theta}(0) \right) - g(\bm{x}_b) } \\
        {\mathcal{N}u\left(\bm{x}_r,\bm{\theta}(0) \right)}
    \end{bmatrix}. \label{eq:NTK_final}
\end{align}

Some remarks about NTKs and the information they provide according to Eq.(\ref{eq:NTK_final}) are in order:
\begin{itemize}
    \item Given that $\bm K(0)$ is a PSD matrix with non-negative eigenvalues $\lambda_i\ge 0$, the training loss decays monotonically with time $t$ and converges to zero in the limit of $t \rightarrow \infty$.
    \item The $i$-th component of the training error decays at a rate proportional to $e^{- \lambda_i t}$. As a result, severe eigenvalue disparity between different loss terms would lead to domination of one term and, therefore, difficulty in training.
    \item It has been shown that larger eigenvalues correspond to low-frequency solutions. This would result in the 'spectral bias' of the PINNs, meaning that the network learns the low-frequency solutions faster. 
\end{itemize}

In the next section, we shall illustrate the usefulness of NTKs in analyzing the behaviour of PINNs dealing with LAD equations.

\section{Numerical experiments}
\label{sec:numExper}
In this section, through the use of the NTK, we study PINN for the linear advection-diffusion equation (\ref{eq:LDA}) and study the impact of the advection speed/diffusion parameter on the training dynamics of PINNs. Unless otherwise stated, we consider the computational domain as $(x,t): [0,1] \times [0,1]$, the initial condition $u(x,0) = \sin{(2 \pi x)}$ and periodic boundary condition. The LAD has a closed form analytical solution given by,
\begin{align}
    u_{Exact} = \sin \left( 2 \pi (x - at) \right) e^{(-\epsilon a^2 t)}. \label{eq:Exact}
\end{align}
For simplicity, periodicity is imposed through Dirichlet boundary condition and according to the analytical solution (\ref{eq:Exact}).
The network architecture used consists of $3$ hidden layers with $200$ neurons in each layer with $\tanh$ as activation function. The network is initialized using the \textit{Xavier} scheme, and \textit{Adam} optimizer is used to minimize the loss function (\ref{eq:Loss}) with learning rate of $0.001$. Furthermore, we generate $N_r = N_b = 300$ collocation points randomly sampled from the computational domain ($x,t$). 

\subsection{Advection-dominated regime}

\begin{figure}[h]
		\centering
		\includegraphics[width=0.99\textwidth]{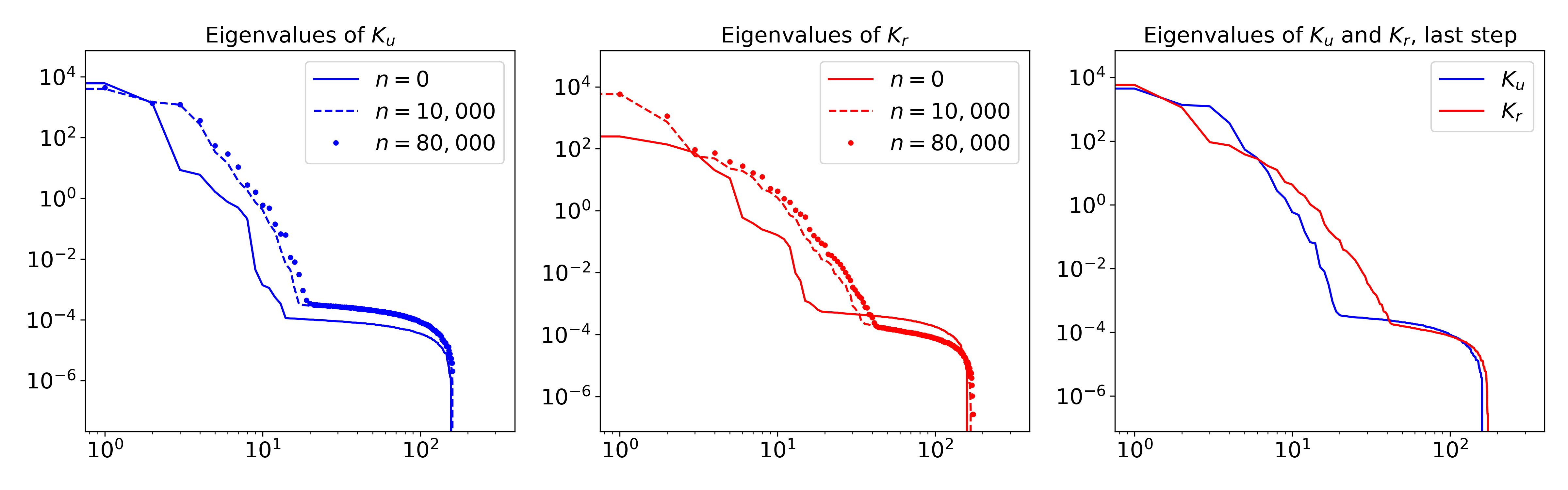}
		\includegraphics[width=0.99\textwidth]{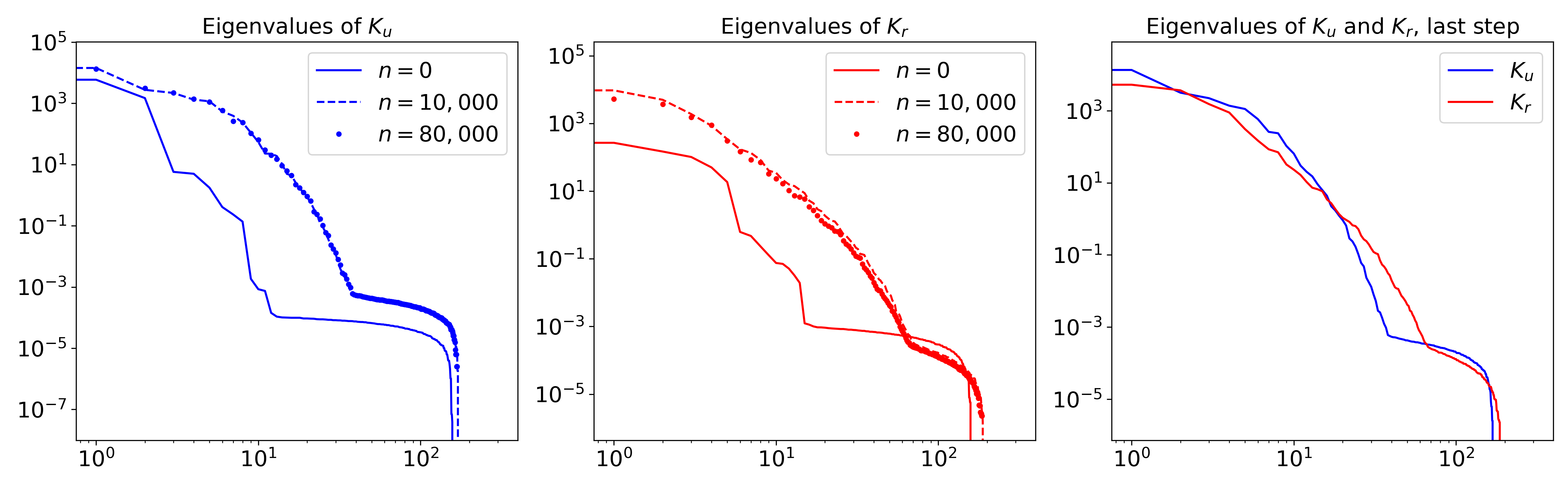}
		\includegraphics[width=0.99\textwidth]{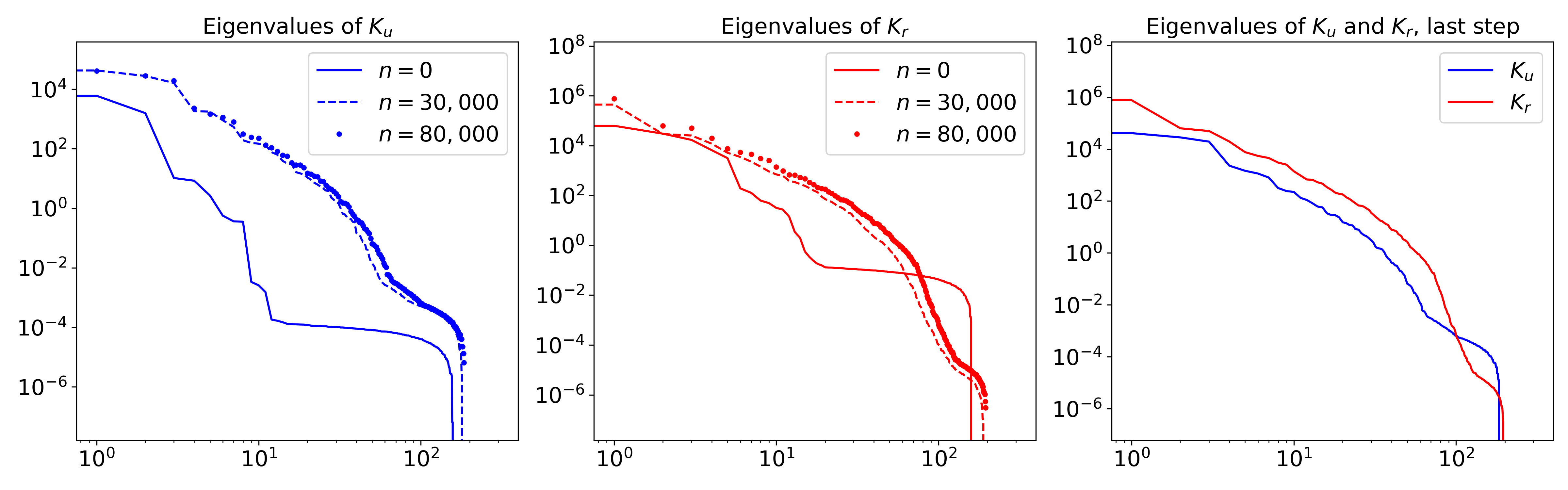}
		\caption{Evolution of NTK eigenvalues for the advection with $a=0.1$ (top), $a=1$ (middle) and $a=15$ (bottom).}
		\label{fig:adv_kernel}
\end{figure}

We first investigate the pure linear advetion equation ($\epsilon=0$) at three different advection speeds, namely $a=0.1, 1$ and $15$. We train the network for $80000$ steps and monitor the eigenvalues of NTKs $\bm{K}_{u}(t)$ and $\bm{K}_{r}(t)$ over the training. Figure~\ref{fig:adv_kernel} illustrates the eigenvalues in descending order for the three different training steps. In all three cases, the figure shows that the eigenvalues decay rapidly to zero such that most of them, except the first few, are near zero. As mentioned in the introduction, the training error decays proportionally to the eigenvalue magnitude and larger eigenvalues correspond to low-frequency eigenvectors. This fact explains that, similar to conventional neural networks, PINNs also tend to learn low-frequency solutions faster and therefore, exhibit 'spectral bias'~\cite{wang2022and, wang2021understanding}. In addition, Fig.~\ref{fig:adv_kernel} shows that all eigenvalues converge, which in turn means that the NTKs converge to deterministic kernels during the training. This is consistent with the NTK theory~\cite{wang2021understanding} discussed in Section~\ref{sec:methodology}.
\begin{figure}[h]
		\centering
		\includegraphics[width=0.99\textwidth]{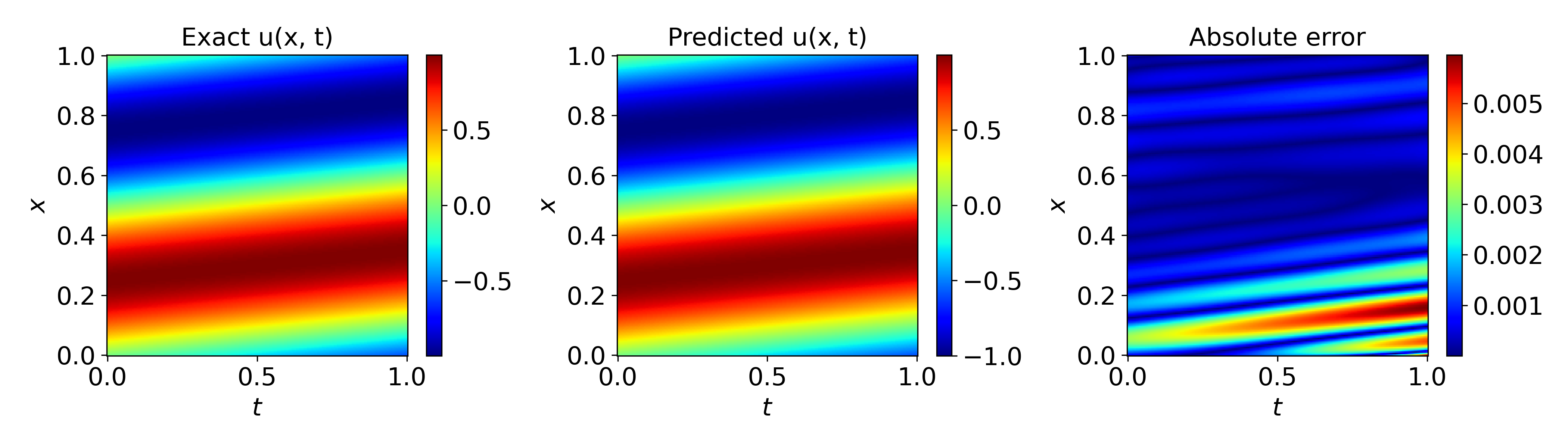}
		\includegraphics[width=0.99\textwidth]{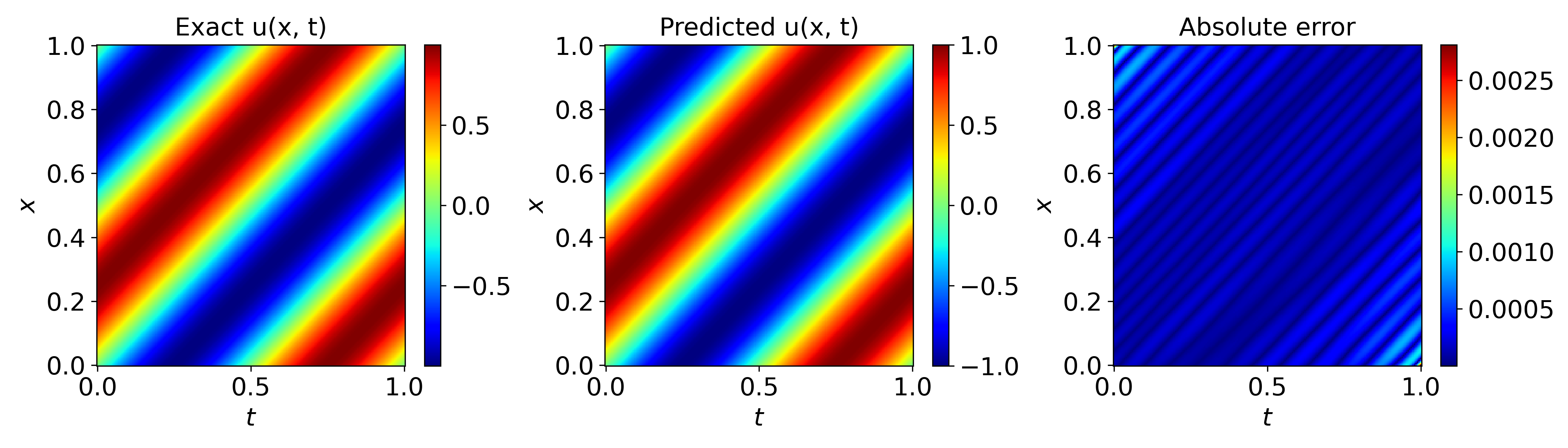}
		\includegraphics[width=0.99\textwidth]{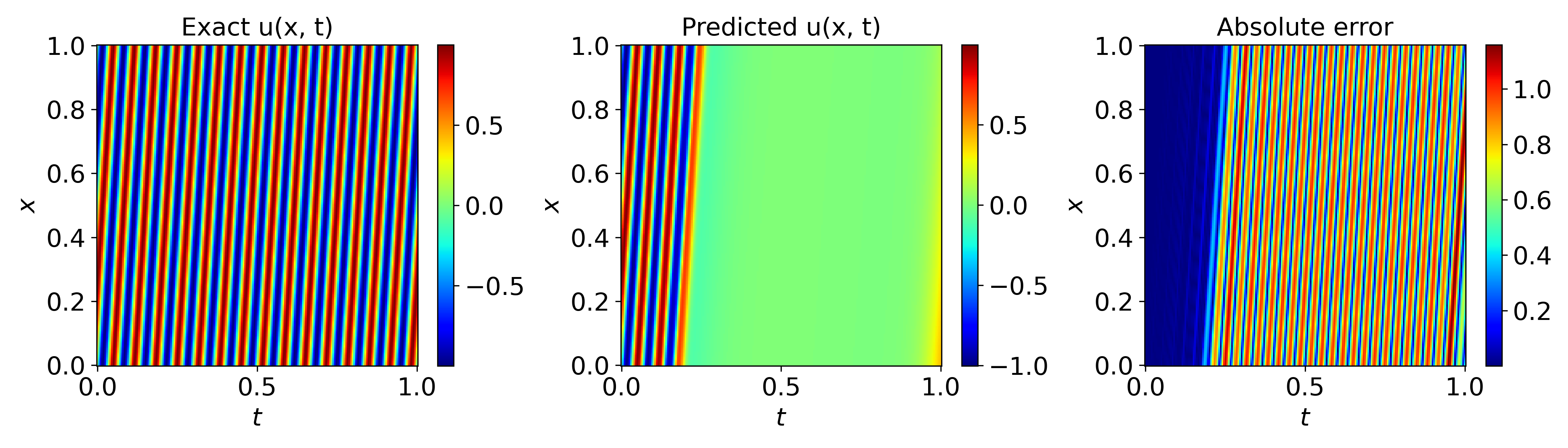}
		\caption{Snapshots of the predicted solution versus exact solution for the advection with $a=0.1$ (top), $a=1$ (middle) and $a=15$ (bottom).}
		\label{fig:adv_result}
\end{figure}

\begin{figure}[h]
		\centering
		\includegraphics[width=0.31\textwidth]{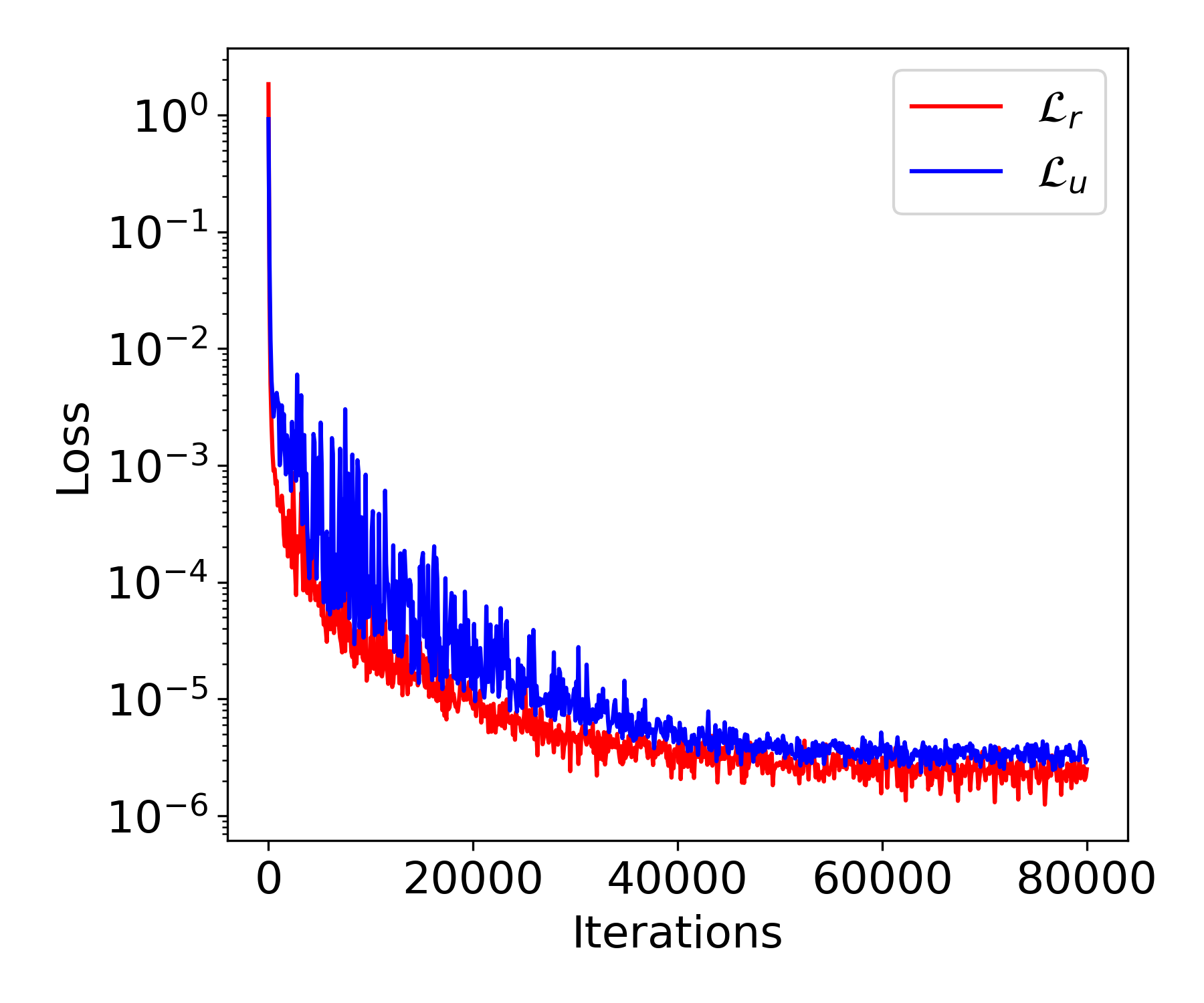}
		\includegraphics[width=0.31\textwidth]{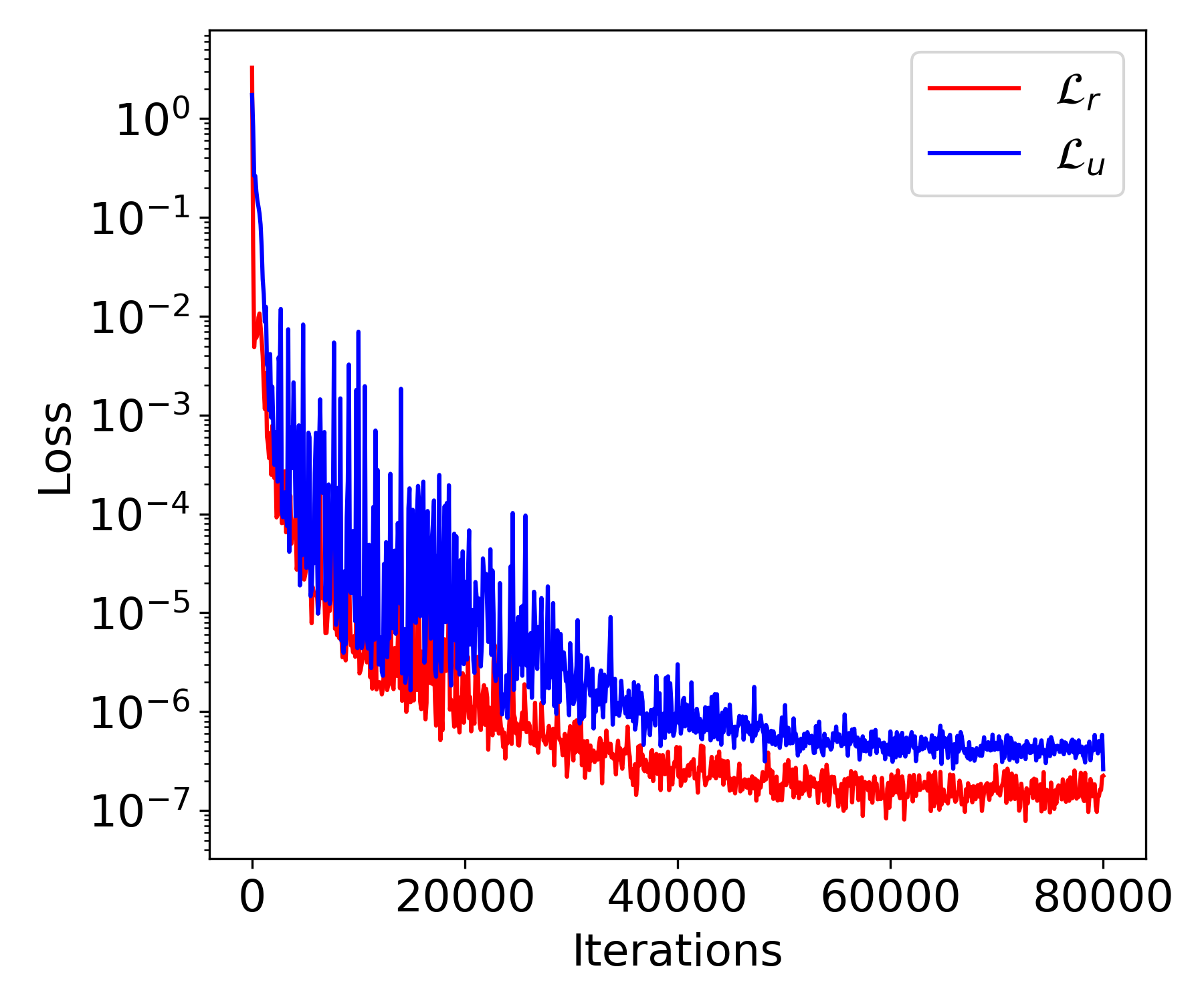}
		\includegraphics[width=0.31\textwidth]{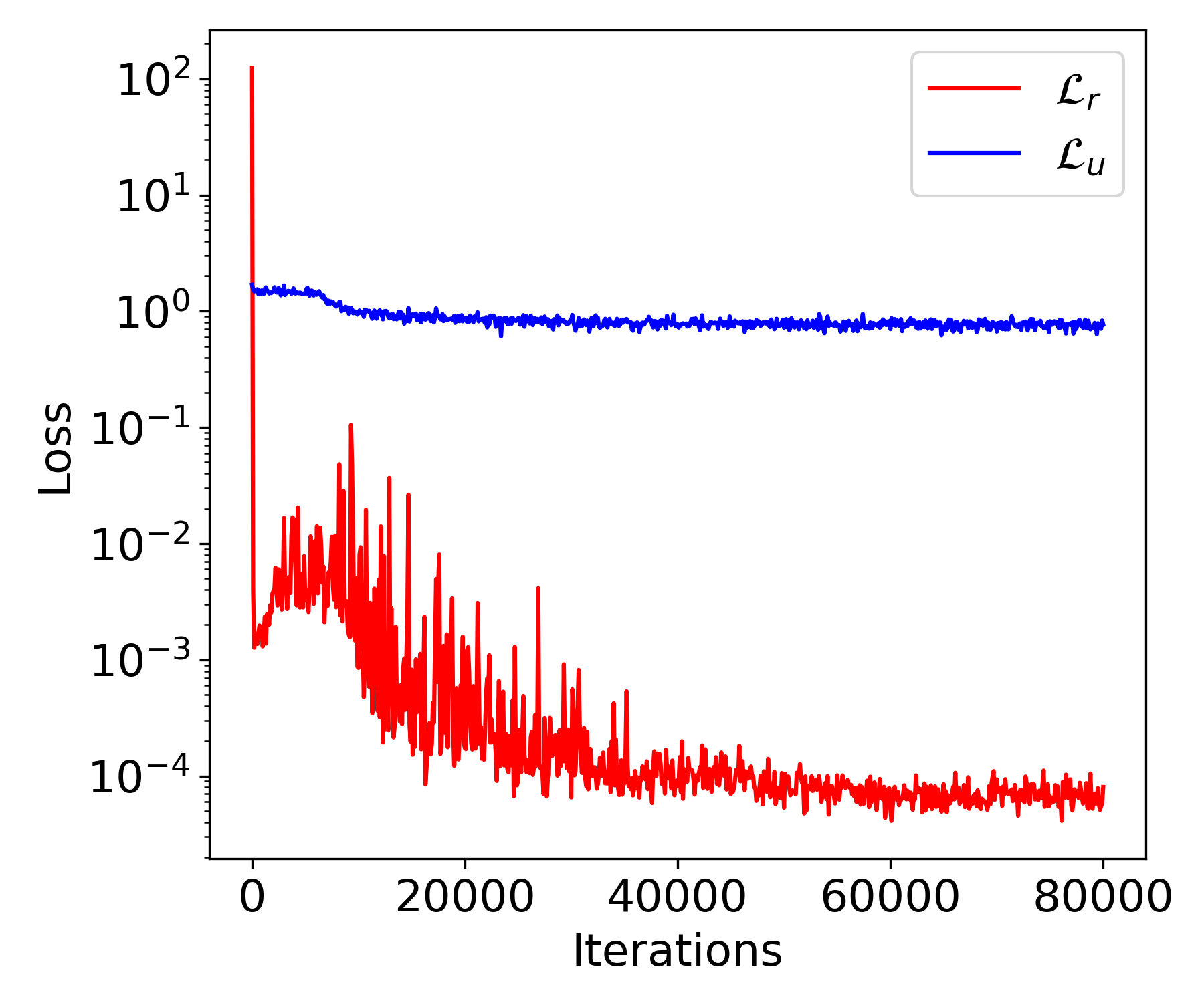}
		\caption{Time history of the different terms in the loss function for the advection with $a=0.1$ (left), $a=1$ (middle) and $a=15$ (right).}
		\label{fig:adv_error}
\end{figure}

Another key thing to note is that, for small values of the advection speed, there is a small disparity between eigenvalues and the predicted solution converges to the exact solution with good accuracy, as shown in Fig. \ref{fig:adv_result}. However, increasing the advection speed to $a=15$ results in a severe disparity between $\bm{K}_{r}$ and $\bm{K}_u$ eigenvalues in a way that $\bm{K}_r$ dominates $\bm{K}_u$ (see Fig. \ref{fig:adv_kernel}). This results in faster decay of residual error $\mathcal{L}_r$ compared to initial/boundary condition error $\mathcal{L}_b$, as illustrated in Fig. \ref{fig:adv_error}. In spite of this fast convergence, the solution converges to a trivial non-physical solution after some time, as it can be clearly seen from Fig.~\ref{fig:adv_result}. Domination of $\bm{K}_r$, indeed, makes the problem ill-posed, as it prevents the network to learn the initial/boundary conditions, leading to trivial constant solution. 

\begin{figure}[h]
		\centering
		\includegraphics[width=0.99\textwidth]{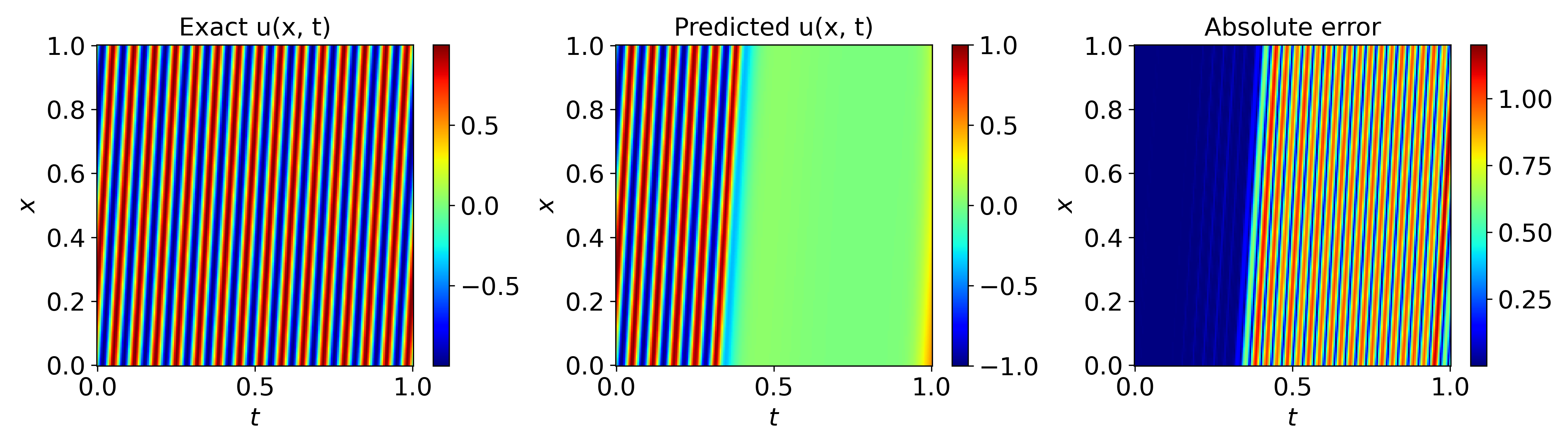}
		\includegraphics[width=0.99\textwidth]{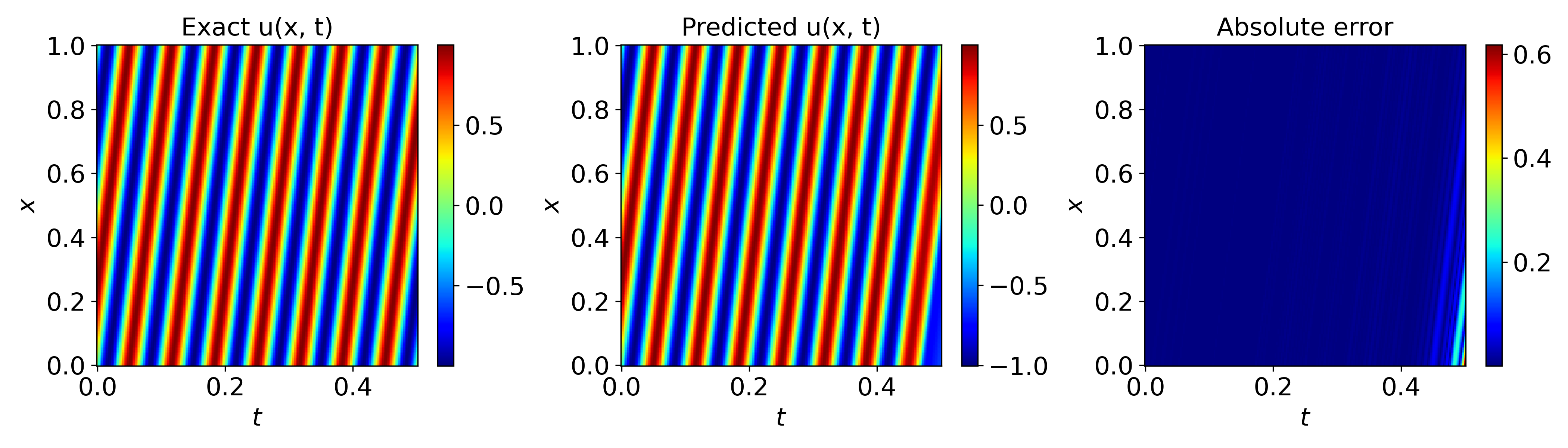}
		\caption{Snapshots of the predicted solution versus exact solution for the advection with $a=15$ using adaptive weights strategy (top) and sub-domain learning on $t=[0,0.5]$ (bottom).}
		\label{fig:adv_strategy}
\end{figure}

To study the importance of properly learning the initial/boundary conditions and the effect of eigenvalue disparity on that, we repeat the training for the case with $a=15$ using the adaptive weights training proposed by Wang et al. \cite{wang2022and}. The idea of adaptive weights strategy is to assign proper weights to different loss components in order to reduce the eigenvalue disparity \cite{wang2022and, wang2021eigenvector}. The results illustrated in Figure~\ref{fig:adv_strategy} demonstrate that while this strategy is, to some extent effective, the predicted solution still converges to trivial solution, albeit at a later time. The failure of PINN, in this case after resolving the eigenvalue disparity issue, can be explained through the lens of the 'spectral bias', since by increasing the advection speed the solution exhibits multi-scale high-frequency behaviour in the space-time domain (see the exact solution in Fig.~\ref{fig:adv_result}). To further investigate this issue, we repeat the training with same setup but on a smaller sub-domain of $(x, t) : [0, 1] \times [0, 0.5]$. This strategy effectively reduces the complexity of the underlying solution and results in robust training of PINN with accurate results. This confirms that spectral bias is an important source of training difficulty. Noteworthy to say that, this is the idea behind the sequence-to-sequence learning \cite{krishnapriyan2021characterizing}, or time adaptive strategies \cite{wang2022respecting, wight2020solving}, which splits the entire space-time domain into multiple subdomains, trains the PINN in sub-domains while marching forward in time. It is important to emphasise that all of these improved learning strategies come at the price of increasing computational and algorithmic complexity. 

\subsubsection{Effect of activation function}
Most of the state-of-the-arts neural network architectures in the literature utilize the rectified linear unit (ReLU) activation function, which have shown success in various deep learning applications. This is, however, not possible in PINNs, as they require sufficiently smooth activation functions, while ReLU is a piecewise linear function incapable of representing second-order derivative and beyond. As such, $\tanh$ is the most widely used activation function for PINN models, despite the fact that it suffers from the vanishing gradient problem and, therefore, is not suitable for deep network architectures. 
\begin{figure}[h]
		\centering
		\includegraphics[width=0.45\textwidth]{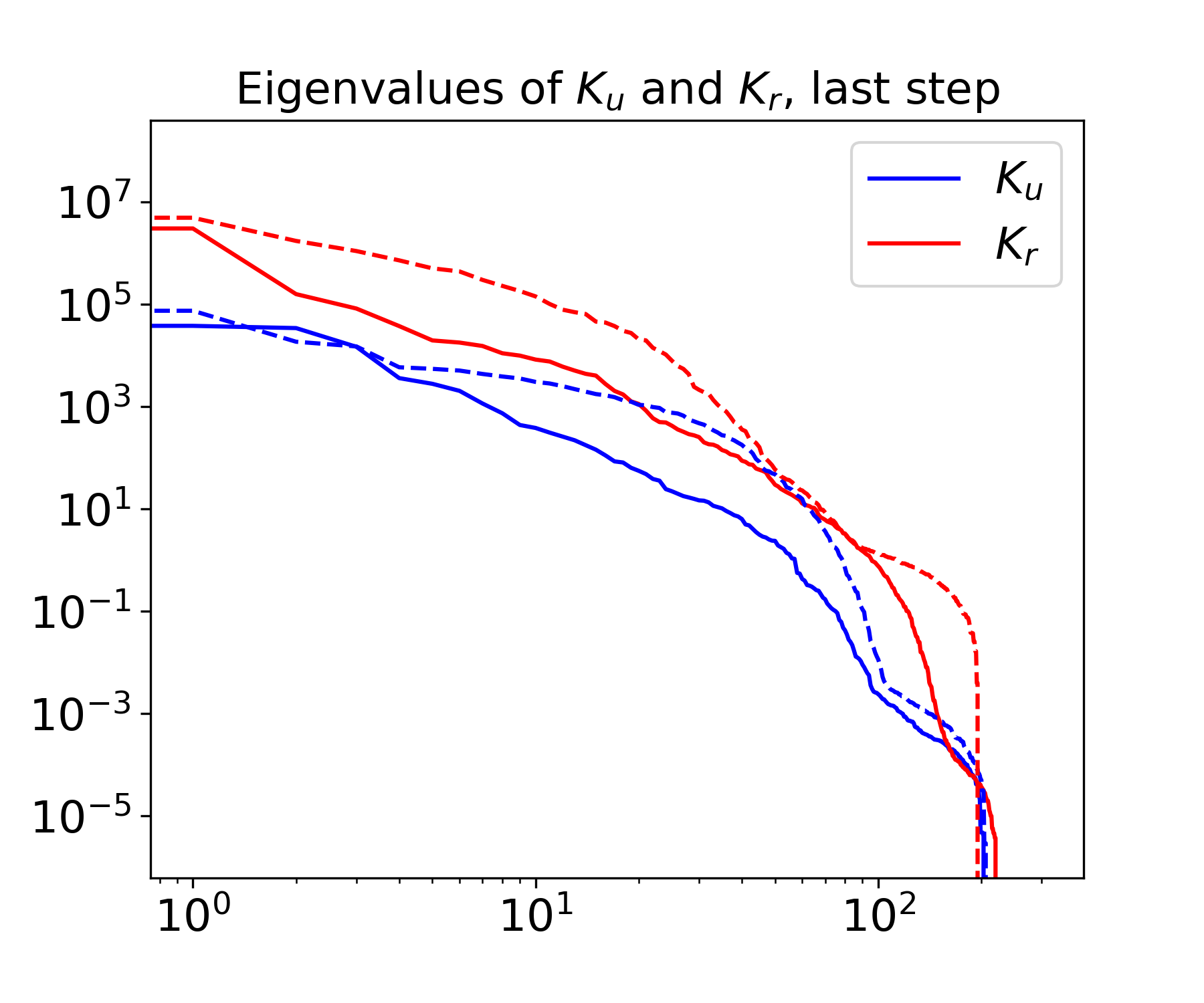}
		\caption{Comparison of NTK eigenvalues for the advection with $a=15$. Solid lines: $\tanh$ activation; dashed lines: $\sin$ activation.}
		\label{fig:kernel_Comp}
\end{figure}

Recently, there has been a growing interest in using periodic functions as an alternative to traditional activation functions \cite{parascandolo2016taming, ramachandran2017searching, sitzmann2020implicit,wong2022learning}. While the first use of periodic activation goes back to three decades ago \cite{sopena1999neural}, it was not until recently that neural networks based on periodic functions were used to solve challenging problems and shown to outperform traditional networks in some applications \cite{ramachandran2017searching, sitzmann2020implicit,wong2022learning}. Noteworthy, is the so-called \textit{SIREN} network which leverages $\sin$ activation function and has been successful in representing complex physical signals \cite{sitzmann2020implicit}. 
\begin{figure}[h]
		\centering
		\includegraphics[width=0.99\textwidth]{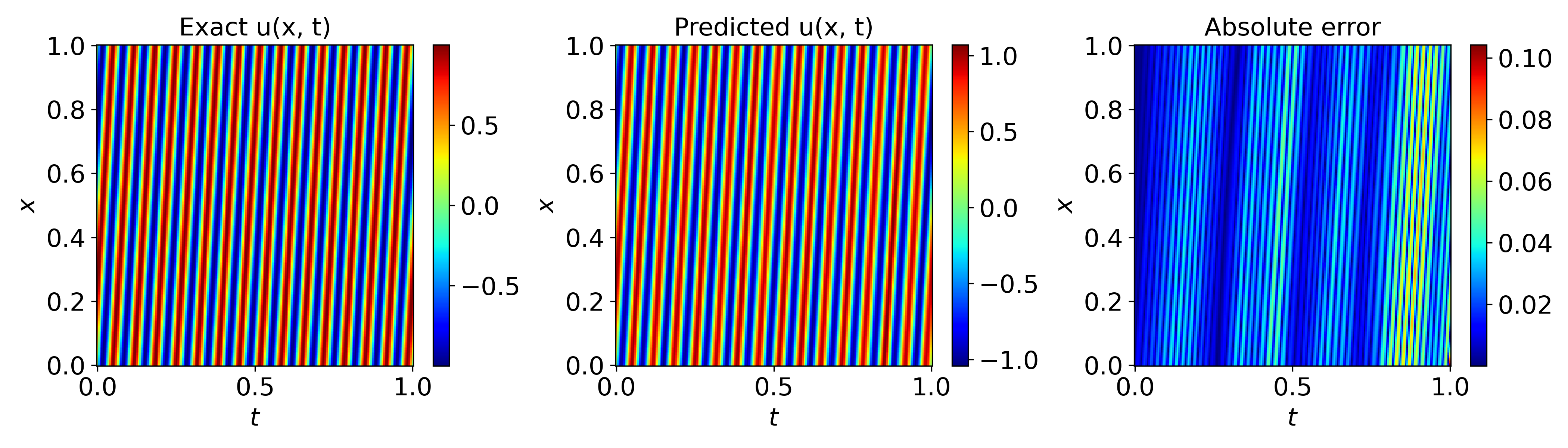}
		\caption{Snapshot of the predicted solution versus exact solution for the advection with $a=15$ using $\sin$ activation function.}
		\label{fig:adv_sin}
\end{figure}

Motivated by this, here we explore the performance of $\sin$ activation function in comparison with the $\tanh$ activation used before by repeating the advection-dominated case with $a=15$. Figure~\ref{fig:kernel_Comp} shows that eigenvalues corresponding to $\sin$ activation decay slower compared to the ones corresponding to $\tanh$. This means that $\sin$ networks might be able to learn higher frequency behaviours than $\tanh$ networks. This is confirmed in Fig.~\ref{fig:adv_sin}, where the predicted solution with $\sin$ network agrees well with the exact solution, while $\tanh$ network fails to learn the correct solution (see Fig.~\ref{fig:adv_strategy} top). Consequently, we can conclude that employing periodic activation functions in PINNs can be an effective approach to address the spectral bias issue without any additional computation overhead or major change in the algorithm (contrary to the previous strategies).

\subsection{Diffusion-dominated regime}
The present section focuses on clarifying the effect of diffusion parameter $\epsilon$ on the training dynamics of PINN. Similar to the previous section, the network is trained for $80000$ steps and for two different parameter values, namely $\epsilon=0.1$ and $10$. 
\begin{figure}[h]
		\centering
		\includegraphics[width=0.99\textwidth]{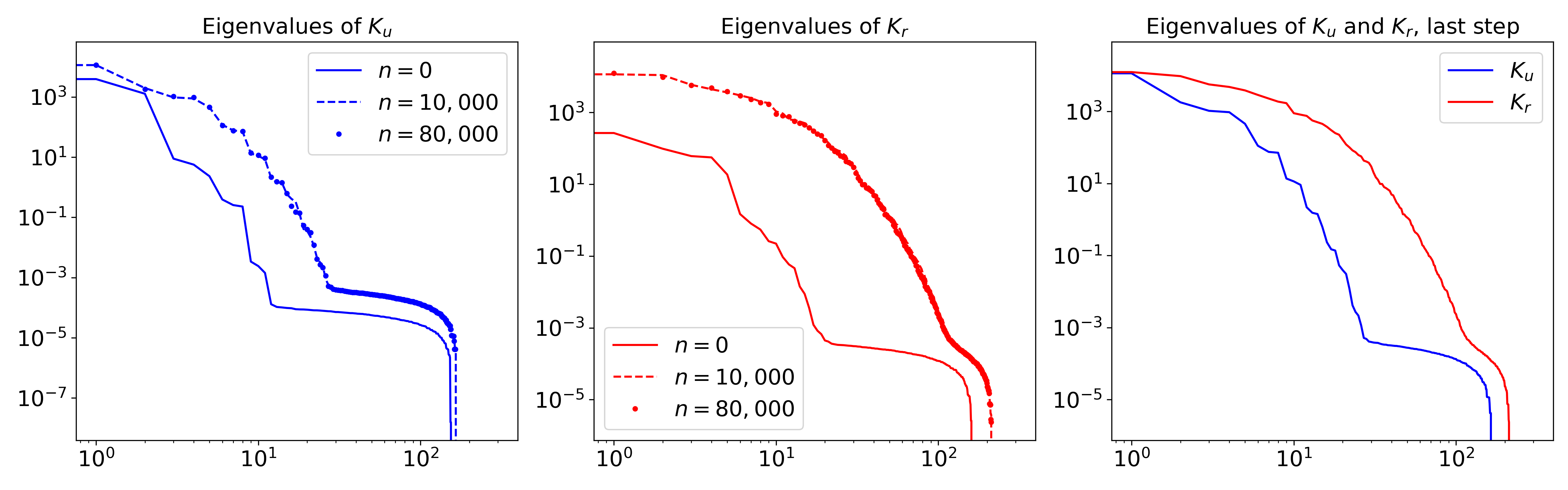}
		\includegraphics[width=0.99\textwidth]{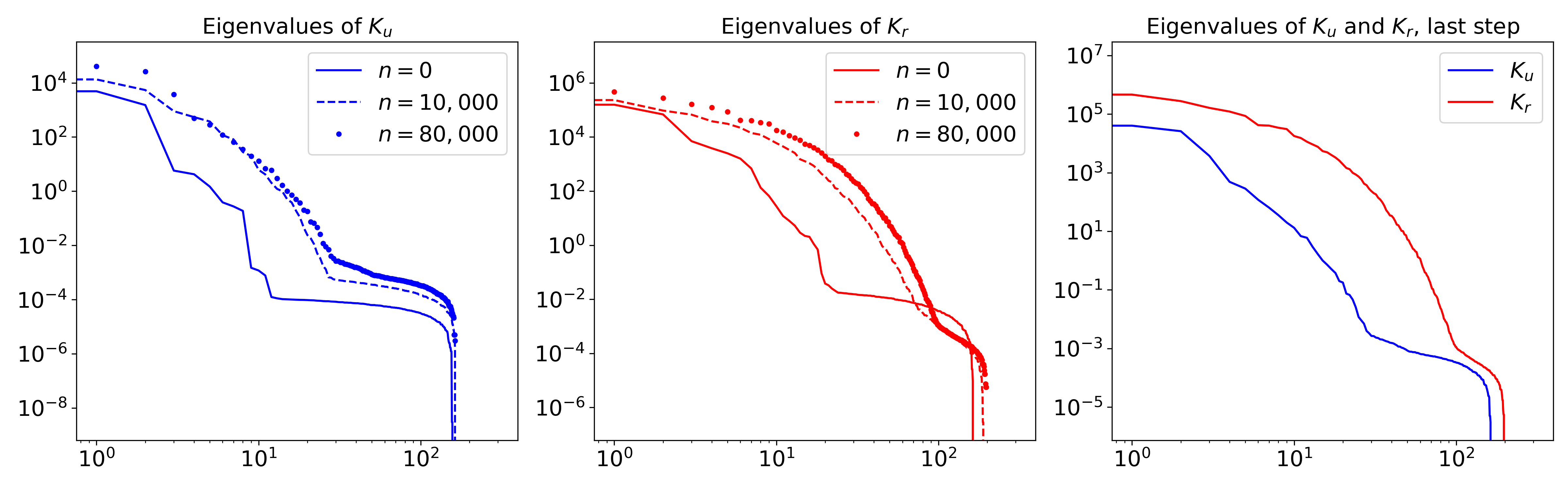}
		\caption{Evolution of NTK eigenvalues for the diffusion with $\epsilon=0.1$ (top), $\epsilon=10$ (bottom).}
		\label{fig:diff_kernels}
\end{figure}
\begin{figure}[h]
		\centering
		\includegraphics[width=0.45\textwidth]{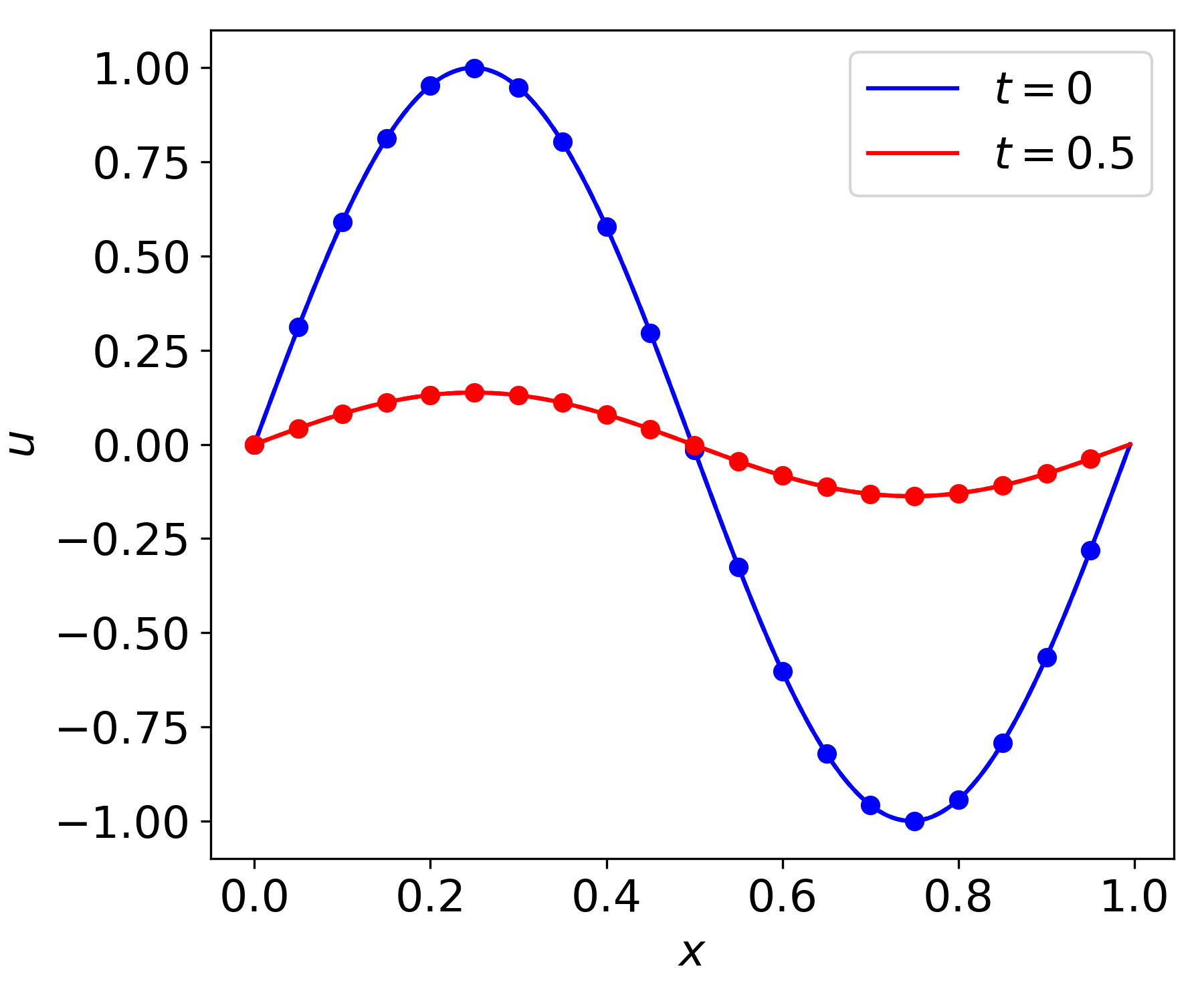}
		\includegraphics[width=0.45\textwidth]{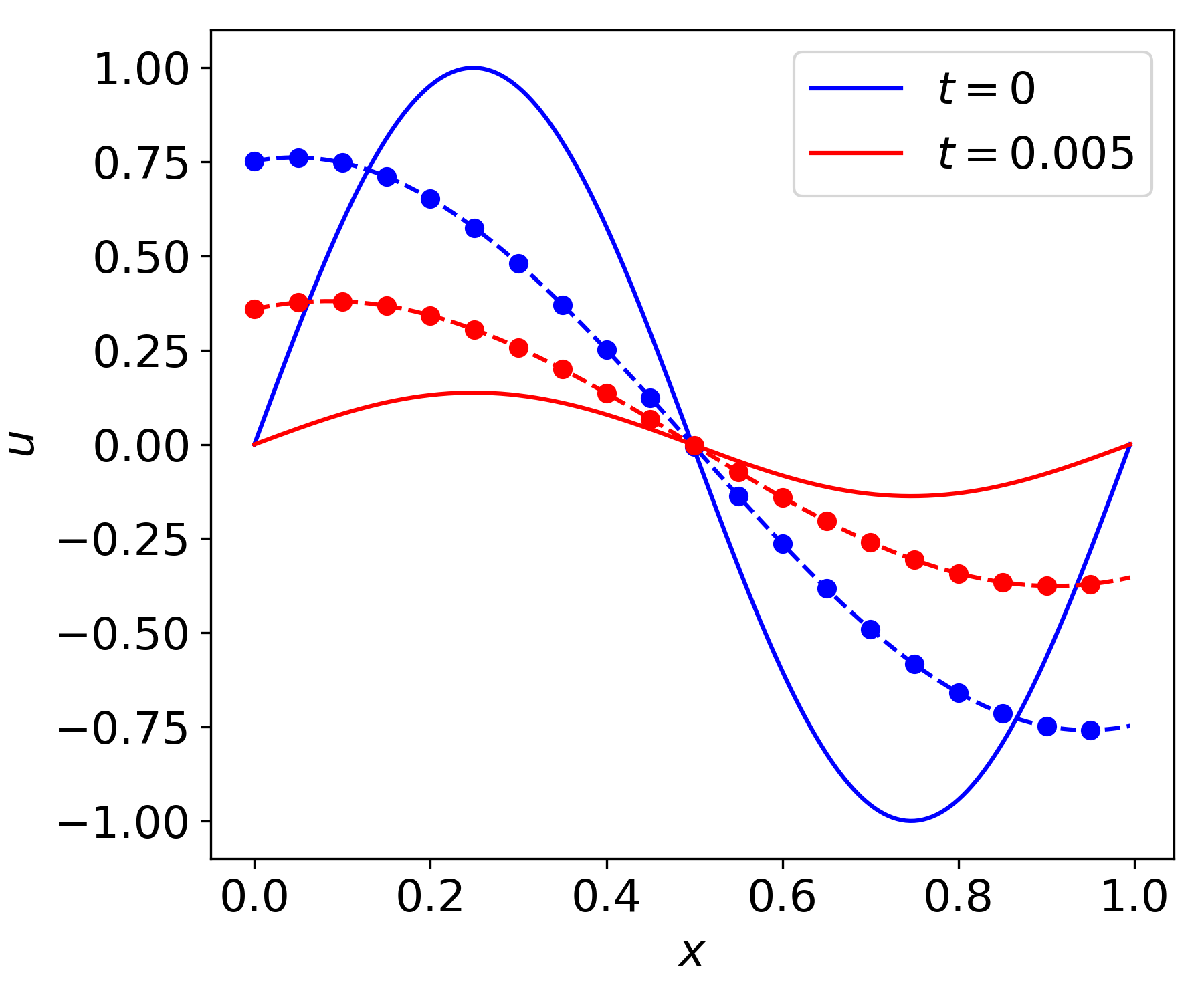}
		\caption{Predicted solution versus exact solution for the diffusion with $\epsilon=0.1$ (left), $\epsilon=10$ (right). Solid lines: exact solution; dashed lines: predicted solution.}
		\label{fig:diff_result}
\end{figure}

\begin{figure}[h]
		\centering
		\includegraphics[width=0.45\textwidth]{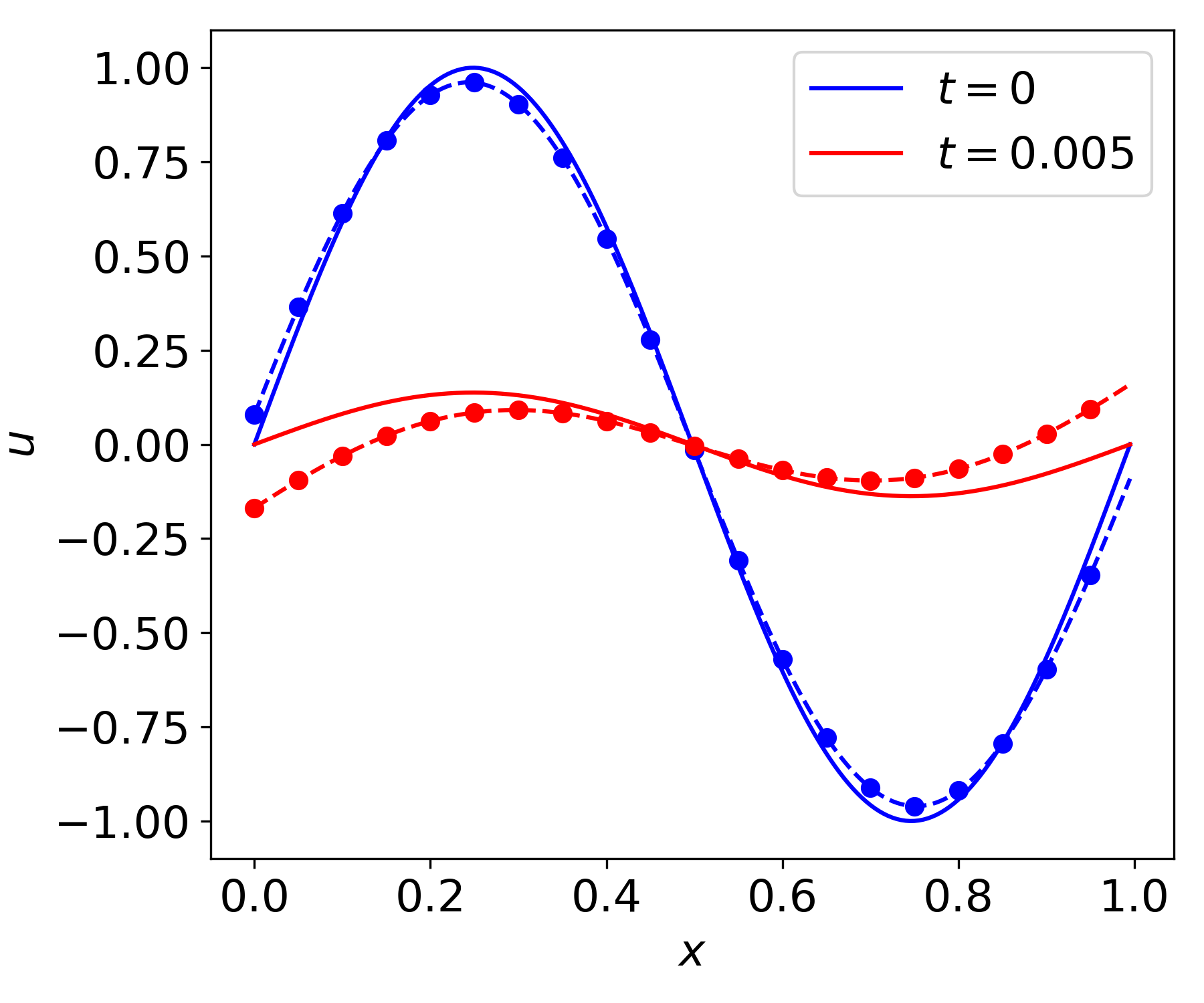}
		\includegraphics[width=0.45\textwidth]{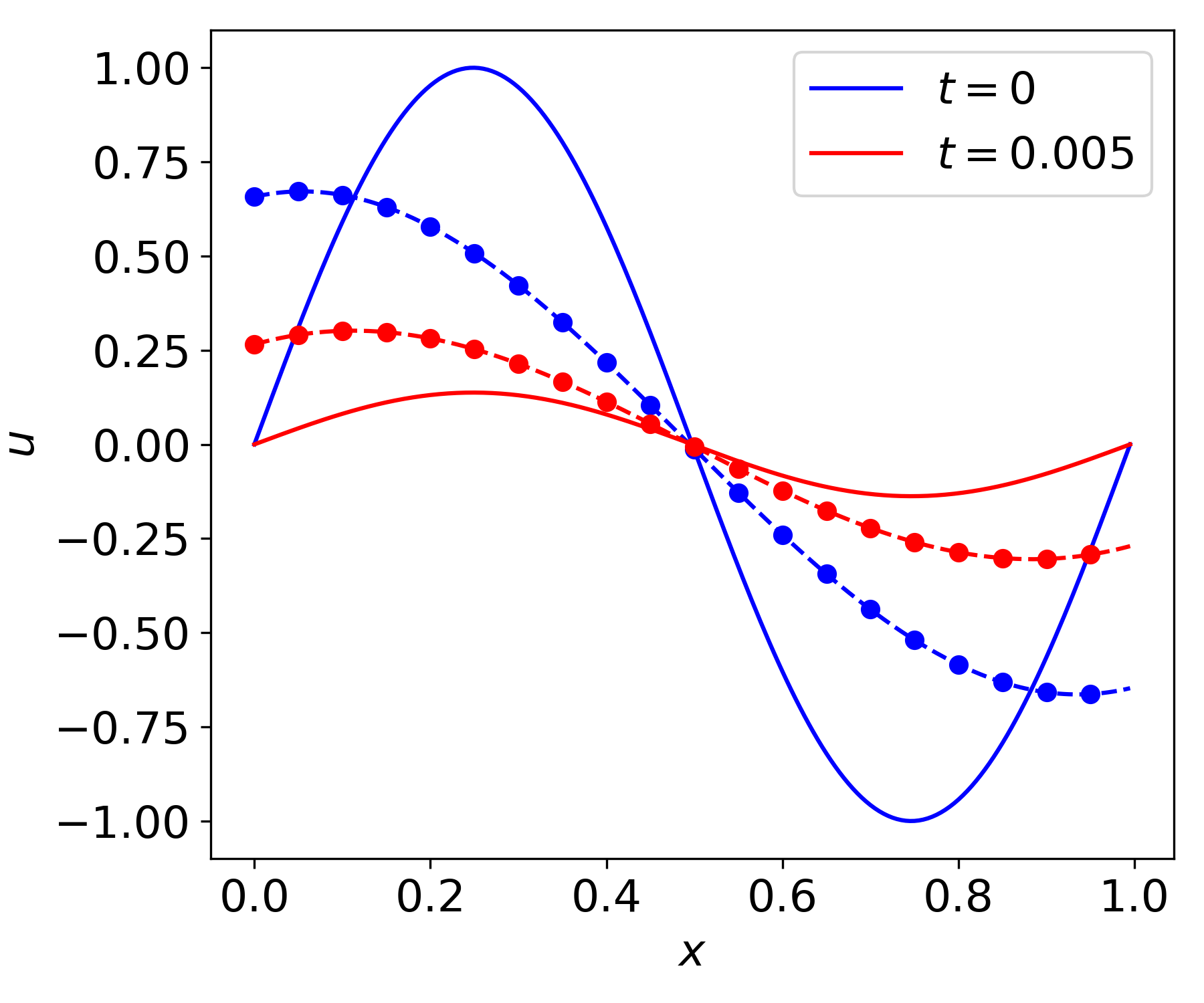}
		\caption{Predicted solution versus exact solution for the diffusion with $\epsilon=10$ using adaptive weights strategy (left) and sub-domain learning on $t=[0,0.1]$ (right) . Solid lines: exact solution; dashed lines: predicted solution.}
		\label{fig:diff_strategy}
\end{figure}

Figure~\ref{fig:diff_kernels} depicts the NTKs for three cases. Qualitatively speaking, NTKs follow similar trends observed in the previous section. As diffusion parameter increases to $\epsilon=10$, the training becomes more difficult, which eventually leads to PINN collapse. However, Fig. \ref{fig:diff_result} shows the underlying exact solution in the case of high diffusivity does not exhibit high-frequency behaviour, as 
the initial condition decays rapidly to trivial solution zero. This counter-intuitive failure of PINN in learning this relatively simple solution of the diffusion equation, could be explained by severe eigenvalue disparity that is observed for this case and is shown in Fig.~\ref{fig:diff_kernels}. Similar to the previous case, the domination of the residual error term prevents the network from learning the initial/boundary conditions. Based on this observation and given the fact that the solution is more or less smooth and trivial in this case, we expect adaptive weights to be a more effective strategy than sub-domain learning or using $\sin$. This is verified in Fig.~\ref{fig:diff_strategy}, where the predicted solution with adaptive weights correctly recovers the exact solution with reasonable accuracy, while the predicted solution even on a very small time domain $t=[0,0.1]$ does not show any significant improvement compared to the baseline PINN results shown in Fig.~\ref{fig:diff_result}. This is in contrast to the advection regime, where sub-domain learning was an effective approach to resolve the training difficulty of PINN.     
\begin{figure}[h]
		\centering
		\includegraphics[width=0.45\textwidth]{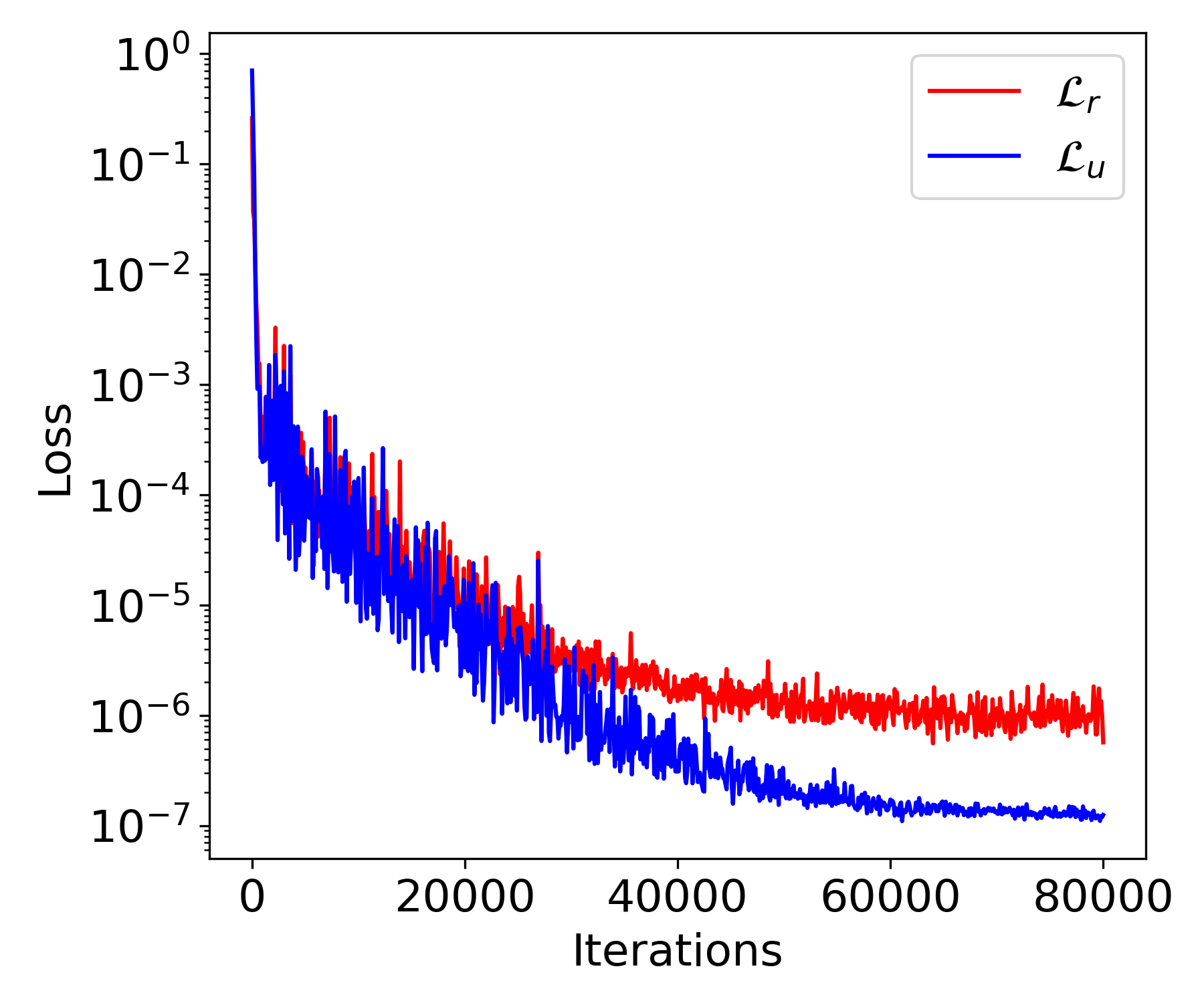}
		\includegraphics[width=0.45\textwidth]{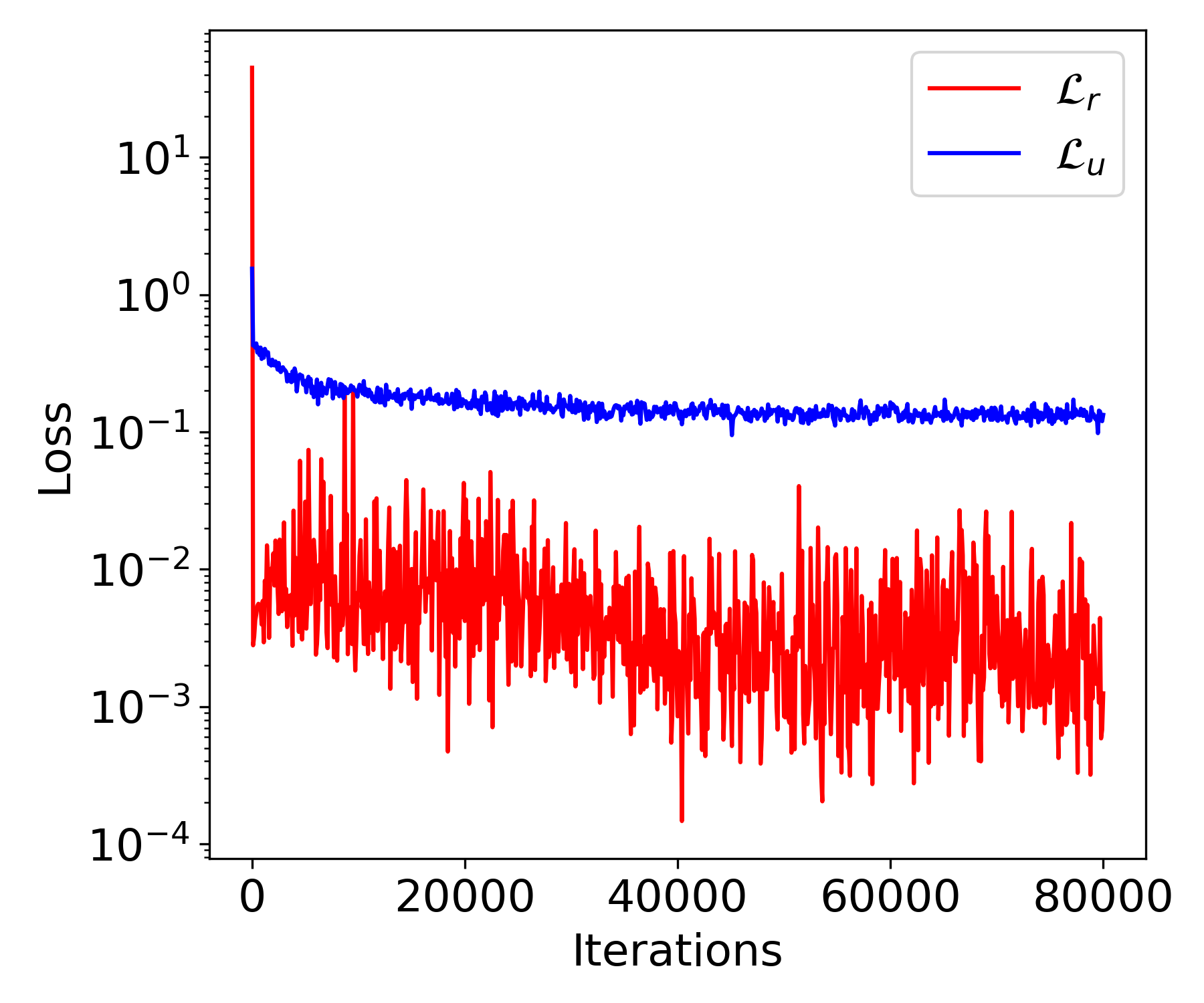}
		\caption{Time history of the different terms in the loss function for the diffusion with $\epsilon=0.1$ (left), $\epsilon=10$ (right).}
		\label{fig:Gauss_Adv1}
\end{figure}

\subsection{Advection-diffusion regime: a numerical example}
Building on the previous sections' findings, we can argue that training failure mechanism of PINNs is different in the advection-dominated and diffusion-dominated regimes. While in both regimes PINNs suffer from the eigenvalue discrepancy, the training difficulty is more prominent in advection-dominated regime due to the underlying high-frequency behaviour of the solution in this regime and the spectral bias of PINNs. A successful PINN model in dealing with advection-diffusion equation (and possibly other class of PDEs) should, therefore, be able to address both eigenvalue disparity and spectral bias, at the same time. 
\begin{figure}[h]
		\centering
		\includegraphics[width=0.99\textwidth]{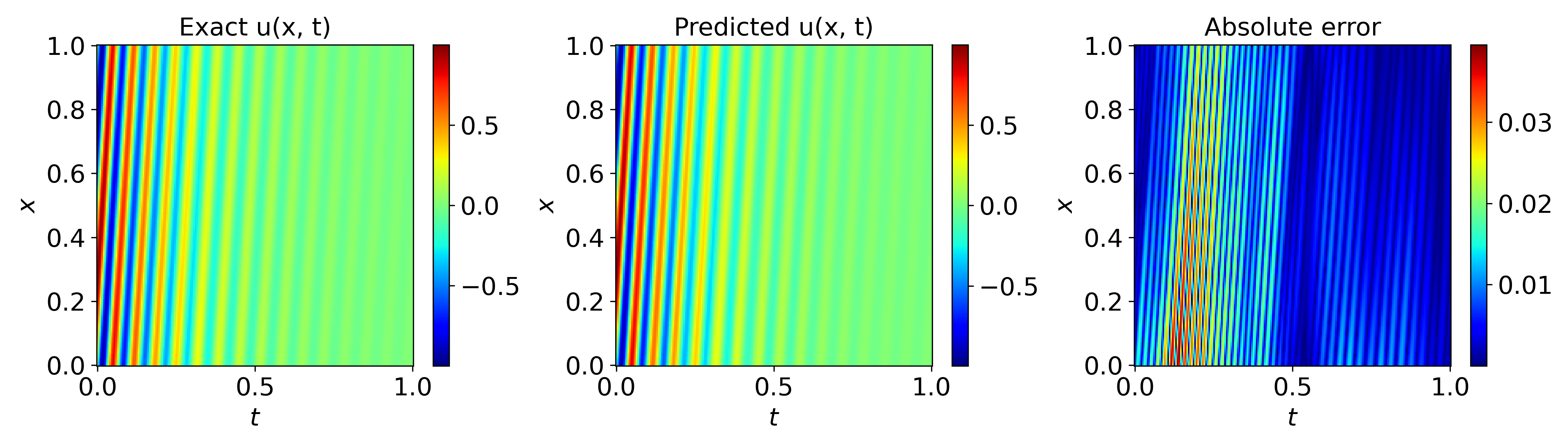}
		\caption{Snapshot of the predicted solution versus exact solution for the advection-diffusion with $a=15$ and $\epsilon=0.1$.}
		\label{fig:advdiff_solution}
\end{figure}

As such, we illustrate the effectiveness of using $\sin$ activation function combined with adaptive weights strategy \cite{wang2021understanding} for the LAD in a difficult regime with $a=15$ and $\epsilon = 0.1$. Figure~ \ref{fig:advdiff_solution} illustrates the good accuracy of the proposed PINN for this case, with the note that the vanilla PINN with $\tanh$ activation \cite{raissi2019physics} is unable to learn the solution.
We shall discuss all our findings in the next Sec. \ref{sec:discussion}

\section{Discussion and Conclusions}
\label{sec:discussion}
In this work, we performed neural tangent kernel analysis of physics-informed neural networks for the linear advection-diffusion equation in advection-dominated and diffusion-dominated regimes. 
Through numerical experiments and eigenvalue analysis of NTKs, we showed that in both regimes the training of PINNs become problematic at large PDE parameters, with possibility of total failure in extreme cases.

Going into more detail, we observed that increasing the advection speed results in: 1) severe eigenvalue disparity between the NTKs of the residual term $\bm{K}_r$ and initial/boundary conditions term $\bm{K}_u$; and 2) the solution in space-time domain exhibits high-frequency behaviour. The former results in domination of the residual term, and thus, it has faster convergence rate through the gradient descent training. This prevents the PINN from learning the correct initial/boundary conditions, which makes the problem ill-posed.  
The latter, on the other hand, poses challenge to PINNs due to the so-called 'spectral bias', a general feature of fully connected neural networks (like the ones used in PINNs) that tend to be biased towards low-frequency solutions. One remedy to the first issue is to reduce the eigenvalue disparity by manipulating the weights of different loss terms as, for example, it is done in the adaptive weights algorithm~\cite{wang2022and}. While using this algorithm improves the results to some extent, other strategies are needed to address the issue stemming from the 'spectral bias'. We showed that splitting the time domain, doing the training on a smaller sub-domain and then marching consecutively in time, as done in the sequence-to-sequence learning \cite{krishnapriyan2021characterizing} or time adaptive re-sampling learning~\cite{wight2020solving}, is an effective approach to resolve this issue. These improved learning strategies, however, increase the computational cost as well as complexity of the algorithm. It was then demonstrated that a simple, yet effective strategy to address the spectral bias issue without any additional computational cost and major change in the algorithm is to use periodic activation functions and in particular $\sin$ activation. NTK analysis supplemented with numerical experiments showed that PINN models with $\sin$ activation are capable of representing higher-frequency solutions compared to $\tanh$ models. Last important thing to note is that, the behaviour of PINNs in advection-dominated regimes is consistent with the behaviour of conventional numerical methods in this regime. It is well know that classical numerical methods are more prone to numerical instabilities in advection-dominated regimes~\cite{suman2017spectral} and there is a long tradition of designing numerical schemes suitable for advection dominated regime~\cite{pirozzoli2011numerical}.

In the diffusion-dominated regime, the eigenvalue analysis of NTKs showed the same trend as in the advection-dominated regime, i.e. increasing the diffusion parameter leads to eigenvalue discrepancy, domination of the residual term in the training process, inability of PINNs in learning the correct initial/boundary conditions and eventually, PINNs failure. However, contrary to the advection-dominated regime, diffusion in general smooths out high-frequency behaviours in the solution. Yet, we showed that for the pure diffusion, where the solution dissipates quickly to a constant steady-state solution and seems to be trivial to learn, the PINN fails and cannot learn the exact solution. This surprising behaviour, however, could be explained by the eigenvalue disparity, which was resolved by employing the adaptive weights learning strategy \cite{wang2022and}. 
It is also interesting to note that, when it comes to diffusion-dominated regimes, the described behaviour of PINNs is in sharp contrast to the behaviour of conventional numerical methods. From numerical point of view, it is in general desirable to have diffusion in our scheme as it dissipates spurious numerical oscillations leading to enhanced numerical stability~\cite{suman2017spectral}. Consequently, one can find a extensive literature on how to optimally add artificial diffusion in order to maintain the stability of a numerical scheme~\cite{pirozzoli2011numerical}.

A unified training strategy that can address both spectral bias and eigenvalue disparity issues is, therefore, in great need. Recent works~\cite{wang2022respecting, wight2020solving} in the literature have shown promising results in that respect. Nevertheless, we showed that using periodic activation like $\sin$ along with adaptive weights \cite{wang2021understanding} is one possibility, which needs to be further investigated for more general and challenging PDEs with non-linearity. Analysing the performance of other periodic activation functions \cite{ramachandran2017searching} in PINNs could be another line of research.

\begin{ack}
This work was supported by the Swiss Federal Office of Energy: “IMAGE - Intelligent Maintenance of Transmission Grid Assets” (Project Nr. SI/502073-01).
\end{ack}

\bibliographystyle{unsrt}
\bibliography{main}

\begin{thebibliography}{10}

\bibitem{yu2018deep}
Bing Yu et~al.
\newblock The deep ritz method: a deep learning-based numerical algorithm for
  solving variational problems.
\newblock {\em Communications in Mathematics and Statistics}, 6(1):1--12, 2018.

\bibitem{sirignano2018dgm}
Justin Sirignano and Konstantinos Spiliopoulos.
\newblock Dgm: A deep learning algorithm for solving partial differential
  equations.
\newblock {\em Journal of computational physics}, 375:1339--1364, 2018.

\bibitem{raissi2019physics}
Maziar Raissi, Paris Perdikaris, and George~E Karniadakis.
\newblock Physics-informed neural networks: A deep learning framework for
  solving forward and inverse problems involving nonlinear partial differential
  equations.
\newblock {\em Journal of Computational physics}, 378:686--707, 2019.

\bibitem{han2018solving}
Jiequn Han, Arnulf Jentzen, and E~Weinan.
\newblock Solving high-dimensional partial differential equations using deep
  learning.
\newblock {\em Proceedings of the National Academy of Sciences},
  115(34):8505--8510, 2018.

\bibitem{yuan2022pinn}
Lei Yuan, Yi-Qing Ni, Xiang-Yun Deng, and Shuo Hao.
\newblock A-pinn: Auxiliary physics informed neural networks for forward and
  inverse problems of nonlinear integro-differential equations.
\newblock {\em Journal of Computational Physics}, 462:111260, 2022.

\bibitem{lou2021physics}
Qin Lou, Xuhui Meng, and George~Em Karniadakis.
\newblock Physics-informed neural networks for solving forward and inverse flow
  problems via the boltzmann-bgk formulation.
\newblock {\em Journal of Computational Physics}, 447:110676, 2021.

\bibitem{jagtap2020conservative}
Ameya~D Jagtap, Ehsan Kharazmi, and George~Em Karniadakis.
\newblock Conservative physics-informed neural networks on discrete domains for
  conservation laws: Applications to forward and inverse problems.
\newblock {\em Computer Methods in Applied Mechanics and Engineering},
  365:113028, 2020.

\bibitem{jin2021nsfnets}
Xiaowei Jin, Shengze Cai, Hui Li, and George~Em Karniadakis.
\newblock Nsfnets (navier-stokes flow nets): Physics-informed neural networks
  for the incompressible navier-stokes equations.
\newblock {\em Journal of Computational Physics}, 426:109951, 2021.

\bibitem{gao2022physics}
Han Gao, Matthew~J Zahr, and Jian-Xun Wang.
\newblock Physics-informed graph neural galerkin networks: A unified framework
  for solving pde-governed forward and inverse problems.
\newblock {\em Computer Methods in Applied Mechanics and Engineering},
  390:114502, 2022.

\bibitem{kashefi2022physics}
Ali Kashefi and Tapan Mukerji.
\newblock Physics-informed pointnet: A deep learning solver for steady-state
  incompressible flows and thermal fields on multiple sets of irregular
  geometries.
\newblock {\em arXiv preprint arXiv:2202.05476}, 2022.

\bibitem{meng2020ppinn}
Xuhui Meng, Zhen Li, Dongkun Zhang, and George~Em Karniadakis.
\newblock Ppinn: Parareal physics-informed neural network for time-dependent
  pdes.
\newblock {\em Computer Methods in Applied Mechanics and Engineering},
  370:113250, 2020.

\bibitem{raissi2020hidden}
Maziar Raissi, Alireza Yazdani, and George~Em Karniadakis.
\newblock Hidden fluid mechanics: Learning velocity and pressure fields from
  flow visualizations.
\newblock {\em Science}, 367(6481):1026--1030, 2020.

\bibitem{mao2020physics}
Zhiping Mao, Ameya~D Jagtap, and George~Em Karniadakis.
\newblock Physics-informed neural networks for high-speed flows.
\newblock {\em Computer Methods in Applied Mechanics and Engineering},
  360:112789, 2020.

\bibitem{cuomo2022scientific}
Salvatore Cuomo, Vincenzo~Schiano Di~Cola, Fabio Giampaolo, Gianluigi Rozza,
  Maizar Raissi, and Francesco Piccialli.
\newblock Scientific machine learning through physics-informed neural networks:
  Where we are and what's next.
\newblock {\em arXiv preprint arXiv:2201.05624}, 2022.

\bibitem{krishnapriyan2021characterizing}
Aditi Krishnapriyan, Amir Gholami, Shandian Zhe, Robert Kirby, and Michael~W
  Mahoney.
\newblock Characterizing possible failure modes in physics-informed neural
  networks.
\newblock {\em Advances in Neural Information Processing Systems}, 34, 2021.

\bibitem{wang2021understanding}
Sifan Wang, Yujun Teng, and Paris Perdikaris.
\newblock Understanding and mitigating gradient flow pathologies in
  physics-informed neural networks.
\newblock {\em SIAM Journal on Scientific Computing}, 43(5):A3055--A3081, 2021.

\bibitem{mishra2022estimates}
Siddhartha Mishra and Roberto Molinaro.
\newblock Estimates on the generalization error of physics-informed neural
  networks for approximating a class of inverse problems for pdes.
\newblock {\em IMA Journal of Numerical Analysis}, 42(2):981--1022, 2022.

\bibitem{de2021error}
Tim De~Ryck and Siddhartha Mishra.
\newblock Error analysis for physics informed neural networks (pinns)
  approximating kolmogorov pdes.
\newblock {\em arXiv preprint arXiv:2106.14473}, 2021.

\bibitem{wang2022and}
Sifan Wang, Xinling Yu, and Paris Perdikaris.
\newblock When and why pinns fail to train: A neural tangent kernel
  perspective.
\newblock {\em Journal of Computational Physics}, 449:110768, 2022.

\bibitem{jacot2018neural}
Arthur Jacot, Franck Gabriel, and Cl{\'e}ment Hongler.
\newblock Neural tangent kernel: Convergence and generalization in neural
  networks.
\newblock {\em Advances in neural information processing systems}, 31, 2018.

\bibitem{arora2019exact}
Sanjeev Arora, Simon~S Du, Wei Hu, Zhiyuan Li, Russ~R Salakhutdinov, and
  Ruosong Wang.
\newblock On exact computation with an infinitely wide neural net.
\newblock {\em Advances in Neural Information Processing Systems}, 32, 2019.

\bibitem{lee2019wide}
Jaehoon Lee, Lechao Xiao, Samuel Schoenholz, Yasaman Bahri, Roman Novak, Jascha
  Sohl-Dickstein, and Jeffrey Pennington.
\newblock Wide neural networks of any depth evolve as linear models under
  gradient descent.
\newblock {\em Advances in neural information processing systems}, 32, 2019.

\bibitem{rahaman2019spectral}
Nasim Rahaman, Aristide Baratin, Devansh Arpit, Felix Draxler, Min Lin, Fred
  Hamprecht, Yoshua Bengio, and Aaron Courville.
\newblock On the spectral bias of neural networks.
\newblock In {\em International Conference on Machine Learning}, pages
  5301--5310. PMLR, 2019.

\bibitem{xu2019frequency}
Zhi-Qin~John Xu, Yaoyu Zhang, Tao Luo, Yanyang Xiao, and Zheng Ma.
\newblock Frequency principle: Fourier analysis sheds light on deep neural
  networks.
\newblock {\em arXiv preprint arXiv:1901.06523}, 2019.

\bibitem{wang2021eigenvector}
Sifan Wang, Hanwen Wang, and Paris Perdikaris.
\newblock On the eigenvector bias of fourier feature networks: From regression
  to solving multi-scale pdes with physics-informed neural networks.
\newblock {\em Computer Methods in Applied Mechanics and Engineering},
  384:113938, 2021.

\bibitem{wang2022respecting}
Sifan Wang, Shyam Sankaran, and Paris Perdikaris.
\newblock Respecting causality is all you need for training physics-informed
  neural networks.
\newblock {\em arXiv preprint arXiv:2203.07404}, 2022.

\bibitem{wight2020solving}
Colby~L Wight and Jia Zhao.
\newblock Solving allen-cahn and cahn-hilliard equations using the adaptive
  physics informed neural networks.
\newblock {\em arXiv preprint arXiv:2007.04542}, 2020.

\bibitem{parascandolo2016taming}
Giambattista Parascandolo, Heikki Huttunen, and Tuomas Virtanen.
\newblock Taming the waves: sine as activation function in deep neural
  networks.
\newblock 2016.

\bibitem{ramachandran2017searching}
Prajit Ramachandran, Barret Zoph, and Quoc~V Le.
\newblock Searching for activation functions.
\newblock {\em arXiv preprint arXiv:1710.05941}, 2017.

\bibitem{sitzmann2020implicit}
Vincent Sitzmann, Julien Martel, Alexander Bergman, David Lindell, and Gordon
  Wetzstein.
\newblock Implicit neural representations with periodic activation functions.
\newblock {\em Advances in Neural Information Processing Systems},
  33:7462--7473, 2020.

\bibitem{wong2022learning}
Jian~Cheng Wong, Chinchun Ooi, Abhishek Gupta, and Yew-Soon Ong.
\newblock Learning in sinusoidal spaces with physics-informed neural networks.
\newblock {\em IEEE Transactions on Artificial Intelligence}, 2022.

\bibitem{sopena1999neural}
Josep~M Sopena, Enrique Romero, and Rene Alquezar.
\newblock Neural networks with periodic and monotonic activation functions: a
  comparative study in classification problems.
\newblock 1999.

\bibitem{suman2017spectral}
VK~Suman, Tapan~K Sengupta, C~Jyothi~Durga Prasad, K~Surya Mohan, and Deepanshu
  Sanwalia.
\newblock Spectral analysis of finite difference schemes for convection
  diffusion equation.
\newblock {\em Computers \& Fluids}, 150:95--114, 2017.

\bibitem{pirozzoli2011numerical}
Sergio Pirozzoli.
\newblock Numerical methods for high-speed flows.
\newblock {\em Annual review of fluid mechanics}, 43:163--194, 2011.

\end{thebibliography}

\end{document}